\documentclass[12pt]{article}

\usepackage[letterpaper, margin=1in]{geometry}

\usepackage{graphicx}
\usepackage[affil-it]{authblk}
\usepackage{float}
\usepackage{caption}
\usepackage{subcaption}
\usepackage{siunitx}
\usepackage{tikz}
\usepackage{verbatim}
\usepackage{amsmath}
\usepackage{booktabs}
\usepackage{pgfplotstable}
\usepackage{hyperref}
\hypersetup{
    colorlinks=true,
    linkcolor=blue,
    filecolor=magenta,      
    urlcolor=blue,
}

\usetikzlibrary{arrows,snakes,backgrounds,patterns,shapes.geometric,calc}
\newcommand{\paficy}[4]{
\pgfmathsetmacro{\cylinderradius}{#1}
\pgfmathsetmacro{\cylinderheight}{#2}
\pgfmathsetmacro{\aspectratio}{#3}
\pgfmathsetmacro{\rotateangle}{#4}
\draw [rotate=\rotateangle](-\cylinderradius,0) arc (180:360:\cylinderradius*1cm and \cylinderradius*\aspectratio*1cm);
\draw[dashed, rotate=\rotateangle] (-\cylinderradius,0) arc (180:0:\cylinderradius*1cm and \cylinderradius*\aspectratio*1cm);
\draw  [rotate=\rotateangle](0,\cylinderheight) ellipse (\cylinderradius*1cm and \cylinderradius*\aspectratio*1cm);
\draw [rotate=\rotateangle] (-\cylinderradius,0) -- (-\cylinderradius,\cylinderheight);
\draw [rotate=\rotateangle] (\cylinderradius,0) -- (\cylinderradius,\cylinderheight);
}
\tikzset{
    dimen/.style={<->,>=latex,thin,every rectangle node/.style={fill=white,midway,font=\sffamily}}
}

\begin{document}

\title{Numerical Study of Suspension Plasma Spraying\\}
\author[1]{Amirsaman Farrokhpanah\thanks{farrokh@mie.utoronto.ca}}
\author[1]{Thomas Coyle}
\author[1]{Javad Mostaghimi}
\affil[1]{\normalsize Department of Mechanical \& Industrial Engineering, University of Toronto}
\date{January 2017}

Published in: \href{https://link.springer.com/article/10.1007/s11666-016-0502-9}{Journal of Thermal Spray Technology}, 26(1-2), pp.12-36

DOI: \href{https://doi.org/10.1007/s11666-016-0502-9}{10.1007/s11666-016-0502-9}

Direct Access: \href{http://rdcu.be/AdpU}{Full Text}

\let\newpage\relax
\maketitle

\begin{abstract}
\noindent A numerical study of suspension plasma spraying (SPS) is presented in the current work. The liquid suspension jet is replaced with a train of droplets containing the suspension particles injected into the plasma flow. Atomization, evaporation, and melting of different components are considered for droplets and particles as they travel towards the substrate. Effect of different parameters on particle conditions during flight and upon impact on the substrate are investigated. Initially, influence of the torch operating conditions such as inlet flow rate and power are studied. Additionally, effect of injector parameters like injection location, flow rate, and angle are examined. The model used in current study takes high temperature gradients and non-continuum effects into account. Moreover, the important effect of change in physical properties of suspension droplets as a result of evaporation is included in the model. These mainly include variations in heat transfer properties and viscosity. Utilizing this improved model, several test cases have been considered to better evaluate the effect of different parameters on the quality of particles during flight and upon impact on the substrate.
\end{abstract}

{\bf \em Keywords:} suspension viscosity, particle trajectory, droplet breakup, suspension plasma spraying.

\section{Introduction}

Suspension plasma spraying is emerging as a powerful coating technique for depositing high quality thermal barrier coatings.  Fine ceramic powders can be used to create stable suspensions in fluids like water or ethanol which are then injected into a plasma jet. The heat from plasma will evaporate the carrier fluid and eventually melt the solid content. The suspension droplets go through several breakups before evaporation is complete. A detailed understanding of how process operating parameters affect the properties of the deposited coating is of great importance. Physical properties of the suspension, operating conditions of the plasma torch, along with injection parameters, i.e., position, angle, and speed, play important roles in this process.

Various experimental and numerical studies have been performed on suspension plasma spraying (SPS). The numerical and experimental study by Fazilleau et al. \cite{Fazilleau_06} focused on interaction between YSZ-ethanol suspension droplets and gas flow in DC plasma spraying. Suspension was injected at a $60{\circ}$ angle aiming the centre point of the nozzle exit. It was shown that plasma flow’s asymmetry due to liquid injection becomes deteriorated at 15 mm downstream of the nozzle exit. Therefore, after 10 to 15 mm, solvent in the suspension has completely evaporated and the solid contents are uniformly mixed with the plasma gas. Their study also pointed out that gas flow velocity variations induced by voltage fluctuations in 200μs time windows play important role in stimulating droplet breakups. In the second part of their study \cite{Delbos_06}, effect of gas fluctuations on solid particle flights was investigated using a simplified 2D model. Effects of plasma parameters on drag and heat transfer of solid particles was included using correction methods. Their model however assumed constant properties for injected particles. Obtained results suggested an optimum substrate standoff distance of 40-60mm.
Waldbillig et al. \cite{Waldbillig_11} performed an experimental study on the effects of torch nozzle size, power, and plasma gas velocity on SPS YSZ coatings. They concluded that the lowest permeability in coatings can be achieved by using small nozzles and high plasma flow rates, while higher deposition efficiency is a results of lower flow rates with small nozzles.
Jabbari et al. \cite{Jabbari_14} performed a numerical study of Nickel-Ethanol suspension spraying. Similarly, Jadidi et al. \cite{Jadidi_15} have numerically focused on flight and impact of Nickel-Ethanol suspension droplets near the substrate and upon impact. Factors such as particle speeds and trajectories along with the shape and position of the substrate were studied. Their results showed that decreasing the substrate’s standoff location from 60 to 40 mm can increase particles’ temperature at substrate by 6\%.
Rampon et al. \cite{Rampon_08} studied application of SPS in producing solid-oxide fuel cells. Their experiment on spraying YSZ suspension in water reported a shift from mono modal particle size distribution to multimodal with the increase of gas Weber number from 5 to 24. More importantly, they concluded that the suspension breakups and droplet size distributions were affected by the viscosity of the suspensions. Their results showed that at high Weber numbers, Ohnesorge number, or viscosity of the suspensions, can become the dominant factor. It has to be noted though that their calculation of Weber and Ohnesorge numbers did not include change of physical properties during flight, i.e. viscosity of the suspension droplets was assumed to remain constant and equal to its initial value at injection time.
Studies like this indicate that the physical properties of suspension materials can have significant impact on the SPS process. For instance, they can tremendously control the atomization process. Suspension defragmentation is a result of domination of different forces acting on the droplets, i.e. (i) inertia and shear forces resulting from the relative velocities between the droplet and plasma gas, (ii) instabilities induced by surface tension forces resisting increase in the surface area of the droplets, and (iii) viscous forces that act by dissipating the instabilities. Three non-dimensional numbers can be used for evaluating the importance of each force compared to another, Weber number ($We=\rho_g v_{rel}^2 d_p / \sigma_p$), Reynolds number ($Re=\rho_g v_{rel} d_p / \mu_g$ ), and Ohnesorge number ( $Oh=\mu_p / \sqrt{\rho_p \sigma_p d_p }$). $\rho$ is the density, $\mu$ is the dynamic viscosity, $\sigma$ is the surface tension coefficient, $d$ is droplet diameter, and $v_{rel}$is the relative velocity between a droplet and plasma gas. Subscripts $p$ and $g$ refer to properties of suspension droplets/particles and plasma gas, respectively. If any of the physical properties were to change, these non-dimensional numbers will be affected and this results have having different patterns of physical phenomena. 
Weber number plays an important role in determining the breakup regimes of suspension droplets. Its value can vary due to changes in gas density, relative velocity, droplet diameter, and surface tension. As the suspension jet is injected into the torch, the values for gas density and relative velocity will change. Meillot et al. \cite{Meillot_13} study shows that the jet’s Weber has a rapid grow upon entering the plasma flow due to high relative velocity and gas density. This value is then reduced with further penetration. As the jet reaches near the torch centre line, Weber will grow again slowly. Droplet diameters is typically controlled by problem parameters such the injector’s size and geometry. Surface tension of droplets can also influence Weber values. For pure liquids, this value will remain constant as the droplets travel in the domain. For suspensions, however, surface tension values may differ. In the preparation process of the suspension, the addition of the solid content to the solvent typically results in a mixture with surface tension values lower than the solvent’s original surface tension. It is however reasonable to assume that this lower surface tension value will not drastically change with the evaporation of the liquid content \cite{Rampon_08} and hence its effect on Weber number and breakup regimes is neglected in current scope. The viscosity of the suspension however, as already suggested by many researches \cite{Waldbillig_11, Toda_06, Chen_08, Chong_71,Lee_70}, goes through significant changes when the liquid content evaporates. The changing value of viscosity can potentially become a key factor in determining droplets’ breakup regime. While changes in viscosity do not affect Weber values directly, effect of viscosity on breakup according to experimental studies becomes important when $Oh>0.1$ \cite{Ashgriz_11}. As the $Oh$ increases, the critical We also increases which can be estimated for large $Oh$ as \cite{Brodkey_95}
\begin{equation}\label{eq:SPS_1}
We_c={We}_{c,Oh->0}\left(1+1.077 Oh^{1.6} \right)
\end{equation}
The critical $We$ marks the transition between breakup regimes \cite{Ashgriz_11} and theoretically drops will experience breakups for values larger than $We_c$. During the suspension flight, as the liquid content evaporates, the viscosity of the suspension increases. This increase of dynamic viscosity ($\mu_p$) will consequently increase Ohnesorge number ($Oh$). According to correlations like equation \ref{eq:SPS_1}, this leads to increases of the $We_c$. In other words, as the droplets loose liquid solvent, they become more resistance to breakups.
In the current study, different available models for prediction of suspension viscosities are reviewed and more suitable models to fit ceramic suspensions of interest are examined. Moreover, different parameters influencing final estimations of viscosity values such as maximum packing factors are studied. Using a 3-dimensional numerical simulation of a simplified plasma torch, suspension droplets are injected into the plasma flow and are tracked as they go through evaporation, and breakup from injection port to the point of impact on the substrate. Various parameters that influence SPS process are then varied to examine their effects on the final particles that impact on the substrate. The goal is to find range of input parameters that results in the highest quality of particles reaching the substrate, particles that ideally have lost all their moister content and are at a higher temperature. At the injection port, parameters that can vary are injection angle, velocity, and location. The effects of torch power and mass flow rate are also examined. For the suspension material, the proper choice of the viscosity model is examined and tested against available experimental data.

\begin{figure}[H]
        \centering
                \includegraphics[width = 6in]{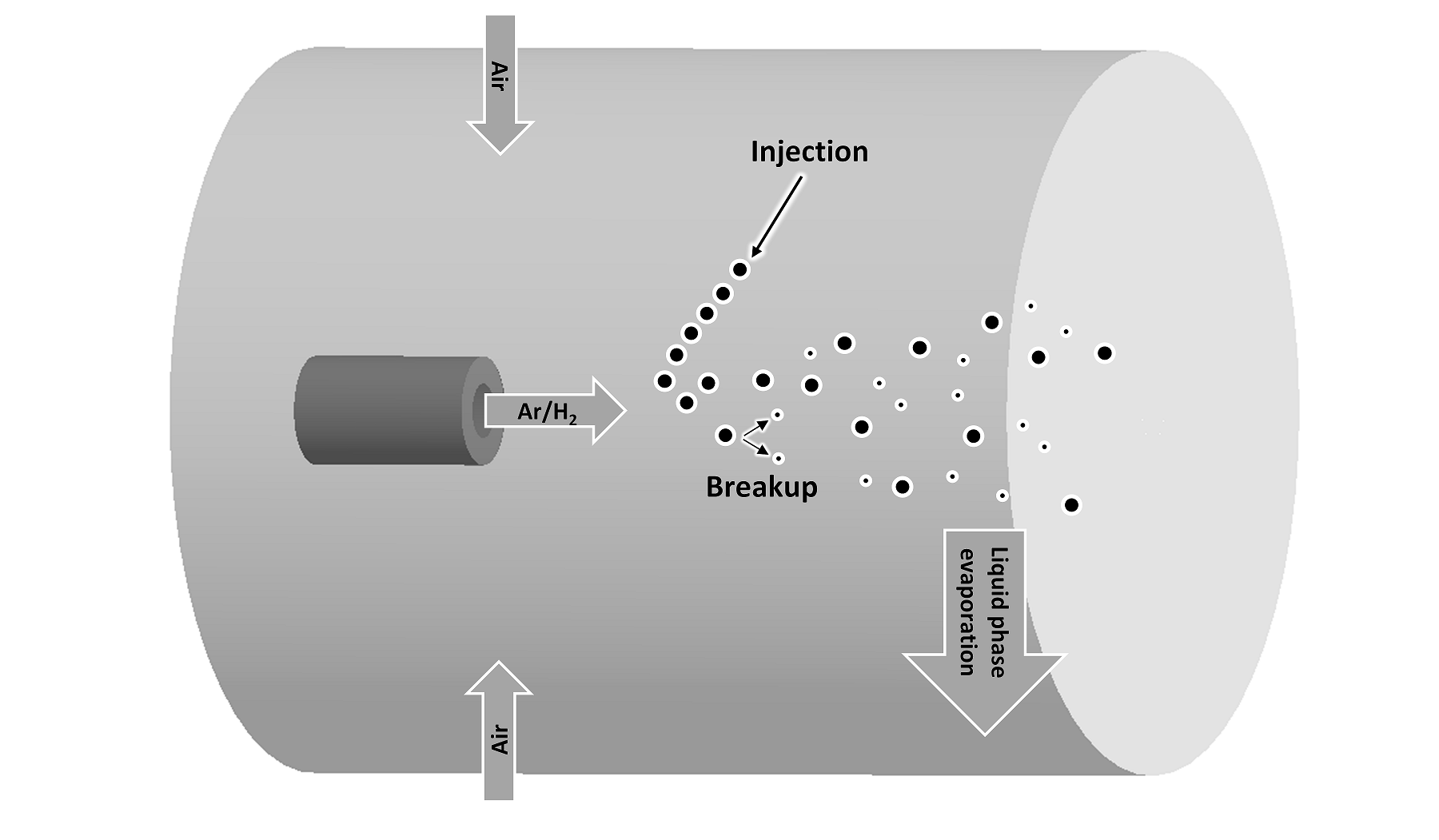}
\caption{Schematic drawing showing torch and the particles that are being injected downstream.}
\label{fig:SPS_1}
\end{figure}

A schematic drawing of the plasma torch and suspension injection used in the current study is shown in \ref{fig:SPS_1}. The plasma torch used in current study has a nozzle exit diameter of 6mm and an anode length of 20mm (3MB plasma spray gun, Orlikon-Metco, Switzerland). The operating conditions used for this torch vary for different test cases and have been summarized in table \ref{tab:SPS_1}. Thermodynamic and physical properties for materials have also been shown in table \ref{tab:SPS_2} \cite{Kang_06,Ansys_13,Wu_09}. Suspension material of interest here is a mixture of Yttria-stabilized zirconia (10\% wt.) and water. The suspension droplets are injected in the domain at a uniform diameter of 150µm. For this study, dispersion of the solid content is neglected and instead, it is assumed that the solid content forms a concentrated ball inside the suspension droplets. Injection site is varied using the axial and radial distances of the injector from the nozzle exit. Suspension particles can also be injected with different initial velocity vectors. For the test cases where the particles are collected on a flat substrate, the standoff distance is 80mm.

\begin{table}[h]
\caption[Summary of operating conditions]{Summary of operating conditions}
\label{tab:SPS_1}
\begin{tabular*}{\textwidth}{l@{\hskip 0.3in \extracolsep{\fill}}c@{\hskip 0.3in}}
 \toprule[1mm]
{\bf Torch geometry and operating conditions}\\
 \hline
 Exit nozzle diameter (mm) & 6\\
Anode length (mm) & 20\\
Current (A) & 450 - 600\\
Voltage (V) & 24.8 - 76.8\\
Thermal efficiency (\%) & 47.1 – 51.2\\
Ar/H2 mass flow inlet (slpm) & 35.4 - 140\\
\toprule[1mm]
{\bf Suspension specifications}\\
 \hline
Ceramic content    &    Yttria-stabilized zirconia (YSZ)\\
Base fluid    &    Water\\
Suspension droplet diameter (μm)    &    150\\
Solid fraction (\%wt)   &    10\\
\toprule[1mm]
\end{tabular*}
\end{table}

\begin{table}[h]
\caption[Materials properties]{Materials properties}
\label{tab:SPS_2}
\begin{tabular*}{\textwidth}{l@{\extracolsep{\fill}}c}
\toprule[1mm]
{\bf YSZ}\\
 \hline
$ \rho (kg⁄m^3)$ &	5560\\
$\mu (Pa∙s)$ &	0.029\\
$\sigma$ (N⁄m)&	0.43\\
$C_p  (J⁄(kg∙K))$&	\shortstack{$1.06343 \times 10^{-6}  T^3-2.188953 \times 10^{-3}  T^2$ \\
				$+1.709671 T+1.466367  \times 10^2, 273<T<873$    \\
				$678.5,T>873$}\\
$\Delta H_{sf}  (kJ⁄kg)$&	710\\
$k (W⁄(m∙K))$&	2.4\\
$T_mp  (K)$&	2975\\
\toprule[1mm]
{\bf Water}\\
 \hline
$\rho (kg⁄m^3 )$ &	998.2\\
$\mu (Pa∙s)$ &	0.001003\\
$\sigma (N⁄m)$ &	0.038 (YSZ-Water mixture)\\
$C_p  (J⁄(kg∙K))$ &	4182\\
$\Delta H_{fg}  (kJ/mol)$ &	Equation \ref{eq:SPS_22}\\
$k (W⁄(m∙K))$&	0.6\\
$T_bp  (K)$ &	373\\
\toprule[1mm]
\end{tabular*}
\end{table}

\section{Numerical methods}

Different numerical methods and treatments have been used in the current study. The Finite Volume based ANSYS Fluent V14.5 (Canonsburg, PA, USA) has been used as the main solver. User defined functions (UDF) written in C programming language are then used for implementing procedures not originally supported by the Fluent solver. The simulation process can be broken into two main stages. Initially, the flow of the plasma gas exiting the nozzle at high speed and temperature is calculated. The discrete phase models are then used to track suspension droplets inside the flow. To avoid numerical complications, instead of injecting a continuous stream of suspension at the beginning, a train of already atomized suspension droplets is injected into the flow. The momentum and heat transfer equations for the gas and particles are coupled in order to capture the effect of particles on the gas flow and vice versa. The evaporation of the liquid phase (here water) inside suspension droplets is reflected in the gas flow by adding species source terms for water vapour. In the following sections, governing equations used in the numerical simulation of the torch and suspension droplets are reviewed.

\subsection{Torch Model}

The plasma torch model utilized here has been schematically shown in Fig. \ref{fig:SPS_1}. The plume is simplified to a jet of fluid entering the domain at high speed and temperature. The torch carries a premixed Ar-H2 gas mixture into atmospheric air. As suspension droplets get injected into the torch, evaporation of liquid phase will also add water vapour to the plasma gas composition. At each cell in the domain, temperature-dependant thermodynamic properties of the plasma gas are calculated based on the mass/volume fraction of each component in this mixture. The chemical reactions and arc influences are neglected. Instead, a volumetric heat source is added inside the torch nozzle. Using torch operating conditions, i.e., current ($I$), voltage ($V$), and thermal efficiency ($\eta$), this volumetric heat source can be estimated as
\begin{equation}\label{eq:SPS_2}
Q = \eta EI/\Omega
\end{equation}
$\Omega$ is the volume of anode. Inclusion of thermal efficiency eliminates the need for adding heat losses at the torch walls and radiation losses. In order to avoid a long torch entrance length, a fully developed turbulent velocity profile in the form of power-law is taken at the Ar-H2 inlet, i.e.,

\begin{equation}\label{eq:SPS_3}
u(r)=u_{max} \left( \frac{r}{R}\right) ^{1/n}
\end{equation}

For the turbulence model, different models and resolutions were tested here. The Realizable k-$\epsilon$ model with the standard wall functions was found to produce the best results at a low computational cost. To avoid deficiencies of the k-$\epsilon$ model in axisymmetric frameworks \cite{Pope_78}, the flow domain has been fully resolved in three dimensions here. Although this eventually adds to the computation time, it was proven to be useful in better capturing trajectories of the suspension droplets injected into the flow in 3D. The extra computational cost is also partially recovered as the k-$\epsilon$ model is computationally less expensive compared to other methods such as RSM \cite{Ansys_13}. The turbulent intensity at the inlet is approximately \cite{Ansys_13}

\begin{equation}\label{eq:SPS_4}
I_t = 0.16 Re_{DH}^{-1/8}
\end{equation}
$Re_{DH}$ is the Reynolds number based on the hydraulic diameter of the torch nozzle. Using this prescribed turbulent intensity, the turbulent kinetic energy is calculated as
\begin{equation}\label{eq:SPS_5}
k=3/2  u_{avg}^2  I_t^2
\end{equation}
and the value of turbulent dissipation rate is evaluated using
\begin{equation}\label{eq:SPS_6}
\epsilon = C_\mu^{3/4} \frac{k^{3/2}}{L}
\end{equation}
$C_\mu=0.09$ is an empirical constant, and L is the length scale approximated to be 0.07 of the hydraulic diameter.

\subsection{Suspension Particle Model}

To avoid numerical complications of primary fragmentation of liquid jets, separate droplets are injected rather than a continuous jet of suspension liquid. The droplets are tracked in a Lagrangian framework as they travel in the domain. The liquid phase in the suspension starts evaporating as the droplets receive heat from the plasma flow. Heat losses due to radiation from the droplets to the surroundings are also included. The plasma plume however is assumed to be optically thin. As the liquid phase vanishes, the concentration of solid contents in the suspension grows. This leads to change in some physical properties, especially viscosity. When the solvent has completely evaporated, the solid contents are tracked until they reach the melting point. After melting, physical properties of particles is the same as that of the molten ceramic. In the following sections, equations and models governing each of these transitions are presented in detail.

\subsection{Momentum}

Suspension droplets/particles here are tracked in Lagrangian manner using a two-way coupled discrete phase model. Plasma gas with the velocity ($u$) accelerates particles inside the domain. Particle acceleration is calculated using a balance between forces acting on the particles and their inertia in the form of \cite{Ansys_13}

\begin{equation}\label{eq:SPS_7}
\frac{du_p}{dt}=F_D (u-u_p ) + F
\end{equation}
Here, $F_D$ is the drag force given as
\begin{equation}\label{eq:SPS_8}
F_D={18\mu}{\rho_p d_p^2}  \frac{C_D Re}{24}
\end{equation}
The drag coefficient $C_D$ is calculated from instantaneous Reynolds number of each particle
\begin{equation}\label{eq:SPS_9}
C_{Df}=\frac{24}{Re}
\end{equation}
for  $Re<0.01$,
\begin{equation}\label{eq:SPS_10}
C_{Df}=\frac{24}{Re} + 3.156 Re^{-\left( 0.18+0.05 log⁡Re \right) } 
\end{equation}
for $0.01<Re<0.2$, and
\begin{equation}\label{eq:SPS_11}
C_{Df}=\frac{24}{Re}+\frac{6}{1+\sqrt{Re}}+0.4 
\end{equation}
for the range $0.2<Re<10^5$ based on empirical correlations for small spherical particle sizes (5-100$\mu$m) \cite{White_06}. For particles in moving in plasma flow, this drag coefficient needs to be modified to take effects of variable properties due to temperature gradient and non-continuum (Knudsen effects) into account. Influence of Basset history term is neglected for particles smaller than 100$\mu$m \cite{Pfender_89}. Effect of variable property is added using proposed method of Lee et al. \cite{Lee_81}. Non-continuum effects also become important for Knudsen number regime $10^{-2}<Kn<1$ \cite{Pfender_89}. This is also considered using correction terms of Chyou et al. \cite{Chyou_89} and Chen et al. \cite{Chen_83}.  A superposition of these corrections, as proposed by Pfender \cite{Pfender_89}, can be used to calculate the final drag coefficient in the form of
\begin{equation}\label{eq:SPS_12}
C_D=C_{Df} f_1 f_2
\end{equation}
where correction factors for strong variation of properties and non-continuum effects are
\begin{equation}\label{eq:SPS_13}
\centering
\begin{gathered}
f_1= \frac{ \rho_\infty \mu_\infty}{\rho_f \mu_f}^{-0.45}
\\
f_2=\left[ 1+\frac{2-a}{a} \frac{\gamma}{1+\gamma} \frac{4}{Pr_f} Kn^* \right]^{-0.45}
\end{gathered}
\end{equation}

The subscript f means properties are calculated at particle’s film temperature, ($T_{cell}+T_p$)/2. $a$ is the thermal accommodation coefficient, $\gamma$ is the heat capacity ratio, and $Pr_f$ is the gas Prandtl number evaluated at film temperature. Here, $Kn^*$ is the Knudsen number based on using effective molecular mean free path ($\lambda_{eff}$) and particle diameter ($d_p$). $\lambda_{eff}$ is calculated from
\begin{equation}\label{eq:SPS_14}
\lambda_{eff}=\frac{2K}{\rho_f V_f C_p } Pr_f 
\end{equation}
with $V_f$, $C_p$, and $K$ being average thermal velocity, specific heat and thermal conductivity, respectively. $F$ in equation \ref{eq:SPS_7} takes thermophoretic effects into account. Thermophoretic forces are exerted on small particles travelling in gas with temperature gradient and are in the opposite direction to this gradient \cite{Ansys_11}. This force is added using proposed method of Talbot et al. \cite{Talbot_80}
\begin{equation}\label{eq:SPS_15}
F=-\frac{6\pi d_p \mu^2 C_s (K^*+C_t Kn)}{ \rho (1+3C_m Kn)(1+2K^*+2C_t Kn)}   \frac{1}{m_p T}  \frac{\partial T}{\partial x} 
\end{equation}
$K^*$ is the ratio of fluid to particle thermal conductivities, $m_p$ is the mass of particle, $C_s$=1.17, $C_t$=2.18, and $C_m$=1.14. Brownian Forces are neglected since there are not many sub-micron particles involved in test cases here. The effects of turbulence on dispersing the particles is included using stochastic tracking method. This method calculates particle trajectories taking instantaneous turbulent velocity fluctuations into account. 

\subsection{Breakup model}

The suspension jet after injection will experience several stages of breakup before reaching the substrate. These breakups need to be included in the numerical method. Melliot et al. \cite{Meillot_11} numerically studied breakup of a train of YSZ-water suspension droplets in the plasma gas. Similarly, Vincent et al. \cite{Vincent_09} used LES-VOF to capture the atomization process of a continuous liquid jet of water injected into the plasma flow. In these simulations, evaporation of liquid phase is neglected. Moreover, the domain of study has been limited to the area close to nozzle exit. Implementation of numerical simulations like these into largescale models, like the one used in the current study, makes solution computationally expensive. Direct simulation of atomization process demands a very fine mesh resolution at fragmentation points and also solving a fluid surface tracking model like VOF \cite{Bussmann_99}. To overcome these issues in the current study, suspension is injected into the domain in the form of droplets. This means that the primary breakup has already taken place. For the secondary breakups and the rest, KHRT (Kelvin-Helmholtz/Rayleigh-Taylor) model [27, 28] is used. This method, which is applicable to high Weber number flows \cite{Ansys_13}, combines effect of aerodynamic forces which create Kelvin-Helmholtz waves along with Rayleigh-Taylor instabilities caused by acceleration of drops into the free stream. These mechanisms capture breakup using drop’s surface wave growth: fastest growing instability will cause the drop to breakup.
For the suspension of study here, the viscosity is updated at each iteration based on drop’s concentration. Breakup model is also influenced by this change of viscosity. Aggregate explosions are neglected here, hence no breakup occurs till all the solvent is evaporated and all solid content has melted. When a molten drop is formed from solid material, breakup is resumed.

\subsection{Heat transfer}

Particle heating and cooling due to convective heat transfer, evaporation, and radiation at particle’s surface is governed by \cite{Ansys_11}

\begin{equation}\label{eq:SPS_16}
m_p C_p  \frac{dT_p}{dt}=h A_p (T_\infty - T_p )+\epsilon A_p \sigma (T_\infty^4-T_p^4 )-\frac{dm_p}{dt} \nabla H_{fg}
\end{equation}

$\sigma$ is the Stephan Boltzmann constant, $\epsilon$ is the emissivity for particles, $m_p$ is the mass of each particle, $C_p$ is the heat capacity, $T_\infty$ is the ambient temperature, $T_p$ is the particle temperature, $h$ is the convective heat transfer coefficient, $dm_p/dt$ is the evaporation rate, and $h_{fg}$ is the latent heat of the liquid phase. The convective heat transfer coefficient is calculated with the help of Nusselt number using Ranz-Marshal correlation
\begin{equation}\label{eq:SPS_17}
Nu_f = 2.0 + 0.6 Re_d^{1/2} Pr^{1/3} 
\end{equation}
$Re_d$ is Reynolds number based on particle diameter and $Pr$ is gas phase Prandtl number. Similar to momentum, effects of temperature gradients and Knudsen number are included using correction factors. Final Nusselt number is corrected using \cite{Chen_83, Chen_83_2}

\begin{equation}\label{eq:SPS_18}
Nu=\frac{h d_p}{k_\infty} =Nu_f f_{h1} f_{h2}
\end{equation}
with $f_{h1} f_{h2} = (f_1 f_2 )^{\frac{1}{0.45}}$. As the solvent inside particles evaporates, heat transfer mechanisms changes. If initial suspension has low ceramic concentration, a shell of molten ceramic might form imprisoning solvent inside \cite{Chen_08}. Also during solvent evaporation, the cloud of ceramic particles may get scattered as a result of sudden explosion. Taking all these effects into account is nearly impossible due to lack of sufficient empirical correlations. Hence, it is reasonable to make simplifying assumptions. Inclusion of explosion and shell formation effects are left for future studies. Instead, it is assumed that evaporation continues till all the liquid content in the droplets is evaporated. Prohibiting scattering of solid particles before they are melted together will possibly result in a larger predictions for final particle sizes.

\subsection{Evaporation of the liquid phase}

During initial stages of droplets’ flight, the liquid phase starts to evaporate. As this phase evaporates, the suspension’s concentration varies significantly as it goes from being dilute to becoming a slurry of moist solid particles that will dry out. The evaporation rate can be corrected to match the drying process of solid-liquid mixtures. Evaporation of the solid content is negligible through all stages. The drying rate of the mixture can be calculated using two different methods. Where experimental data for calibration is available, the model presented by Stendal \cite{Stendal_13} and Seader et al. \cite{Henley_11} for solid-liquid slurries can be used to estimate the evaporation rate

\begin{equation}\label{eq:SPS_19}
\frac{dm_p}{dt}=m_p  \frac{dW}{dt}=m_p  f(W)  exp⁡ \left(-\frac{ \nabla H_{fg}}{R T_p } \right)
\end{equation}
$W$ is the wet basis moisture content, and R the universal gas constant. $f(W)$ is evaluated using experimental data for every particular suspension. Where empirical data for calculation of $f(W)$ is not available, the evaporation and mass transfer rates can be obtained using Sherwood number correlation \cite{Ansys_11}

\begin{equation}\label{eq:SPS_20}
Sh=\frac{k_c d_p}{D} = 2 + 0.6 Re_{DH}^{0.5} Sc^{1/3}
\end{equation}
$D$ is the binary diffusion coefficient, Schmidt number is $Sc=\mu / \rho D$, and $k_c$ is the mass transfer coefficient. By taking an estimate for the binary diffusion coefficient, the mass transfer coefficient becomes available using equation \ref{eq:SPS_20}. Therefore, the evaporation rate can be calculated using
\begin{equation}\label{eq:SPS_21}
\frac{dm_p}{dt}=k_c  M A_p  (C_s-C_b)
\end{equation}
where M is the molecular weight of the liquid content, and $C_s$ and $C_b$ are vapour concentrations at drop’s surface and the bulk of gas, respectively. The latent heat of evaporation here is evaluated at particle temperature using
\begin{equation}\label{eq:SPS_22}
\nabla H_{fg}=-0.0438926 T_p + 57.0735 (kJ/mol)
\end{equation}
for water \cite{Marsh_87} and using
\begin{equation}\label{eq:SPS_23}
\nabla H_{fg}=A~exp⁡ \left( -\alpha \frac{T_p}{T_c} \right) \left(1-\frac{T_p}{T_c} \right)^\beta  (kJ/mol)
\end{equation}
for ethanol from \cite{Majer_85, Acree_10}, with A=50.43 kJ/mol, $T_c$=513.9 K, $\alpha$=-0.4475, and $\beta$=0.4989. As particles go through multiple breakups before the liquid phase is completely evaporated, the solid content of each particle is estimated using the assumption that each breakup divides the solid content in the particle evenly between the forming children based on their diameters.

\subsection{Melting of the solid content}

\begin{figure}[H]
        \centering
                \includegraphics[width = 6in]{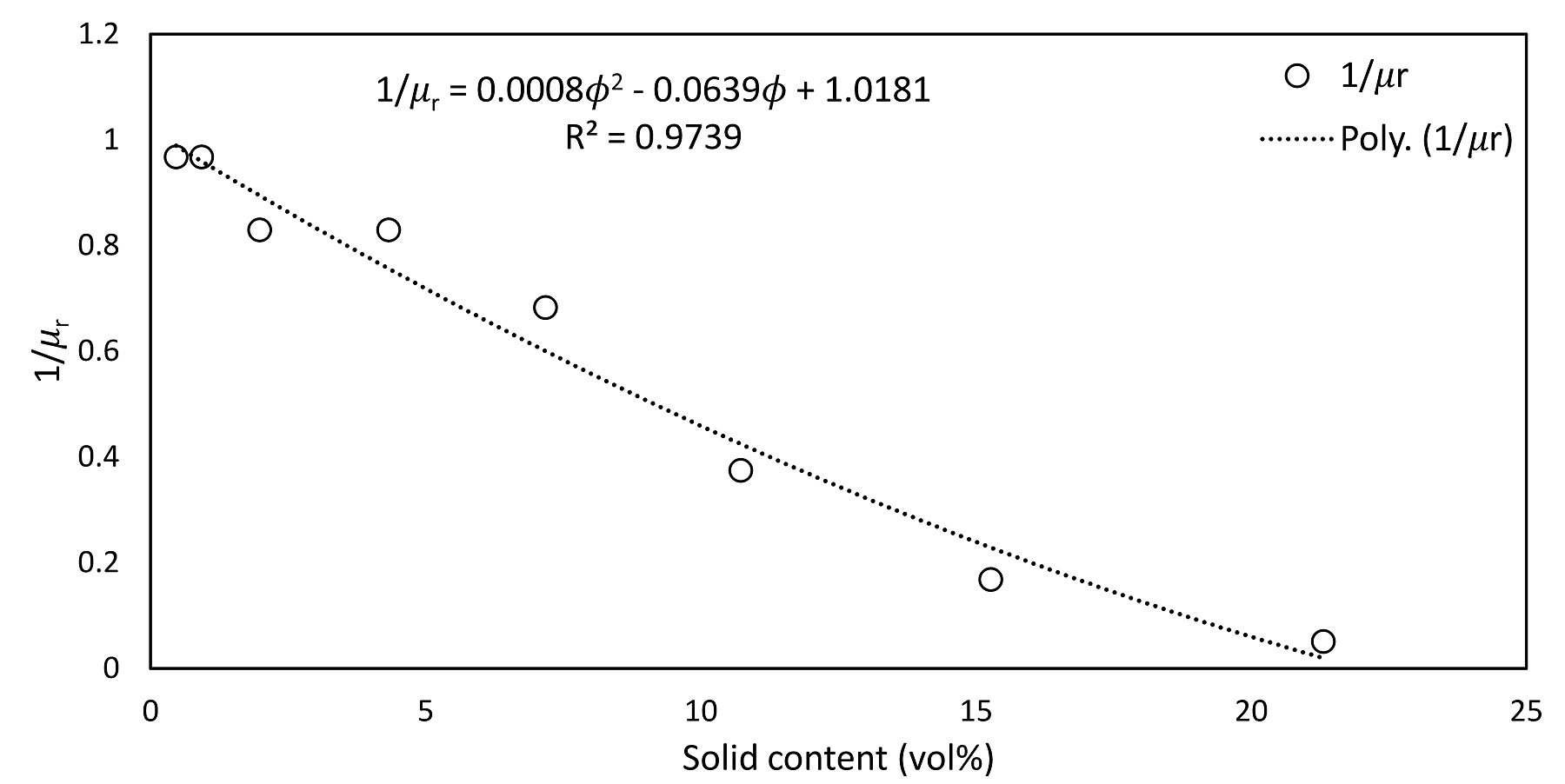}
\caption{Mushy zone effect on viscosity values of YSZ.}
\label{fig:SPS_2}
\end{figure}

After the liquid phase is completely evaporated, the solid content is heated till reaching the melting point of the solid material. $C_p$ values for particles are now only calculated from the properties of the solid material. Viscosity values are also adjusted in the absence of liquid phase. This means that when the fraction of liquid content in each drop becomes negligible, the viscosity for the mixture, which now consists of only solid particles, is set to a large value to mimic non-molten solid matter. This is demonstrated in Fig. \ref{fig:SPS_2} for YSZ. In a 10K window around the melting temperature, the viscosity is linearly dropped from this large value to the value for molten YSZ. This will also be an estimate for the mushy behaviour of the melting process.

\begin{figure}[H]
        \centering
                \includegraphics[width = 6in]{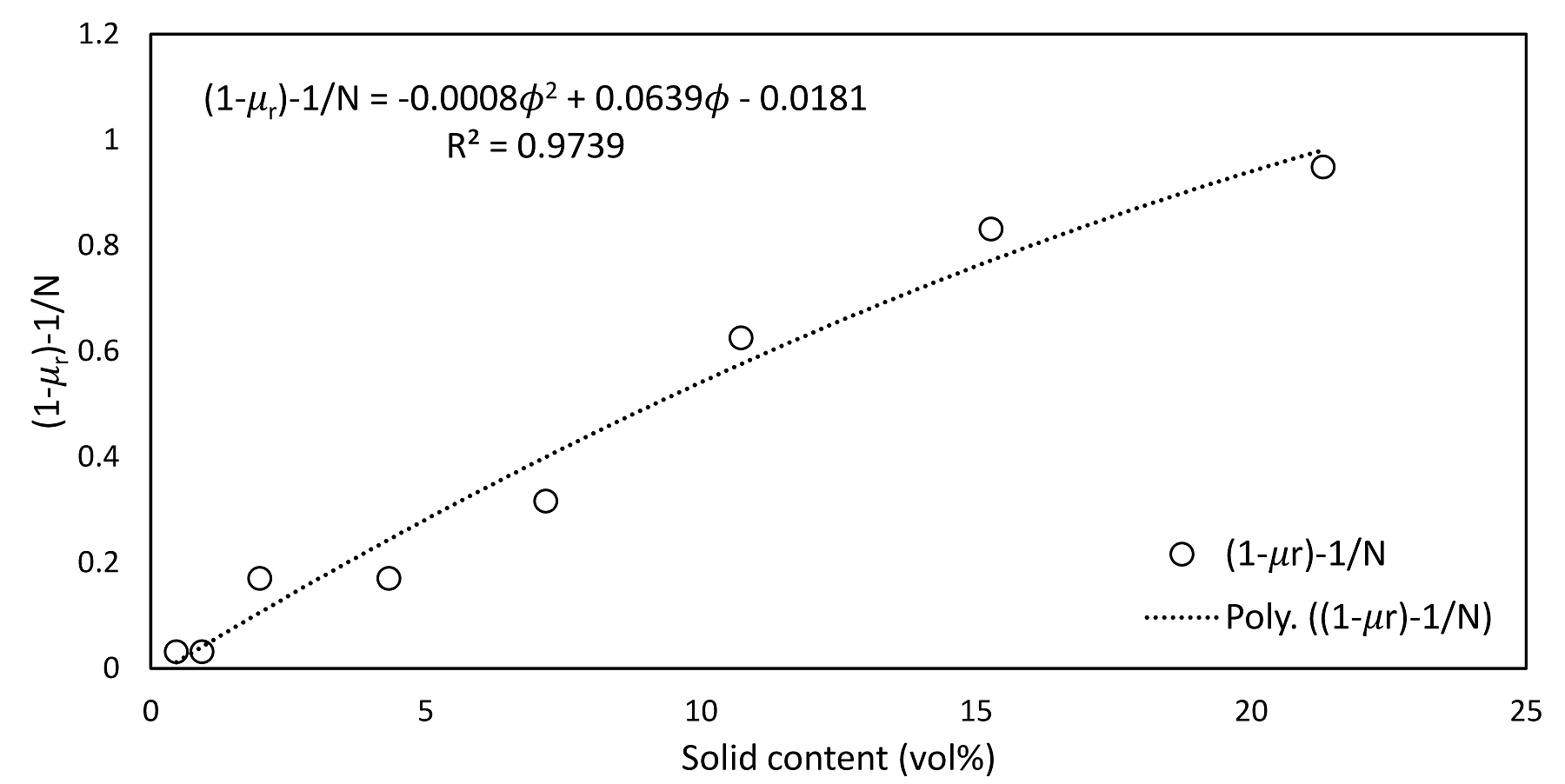}
\caption{Effect of latent heat of melting on $C_p$ values.}
\label{fig:SPS_3}
\end{figure}

The latent heat consumed during melting process is also included by modifying $C_p$ values of YSZ. This is done using 
\begin{equation}\label{eq:SPS_24}
C=
\begin{cases}
    C_s,& \text{if } T \leq T_1\\   
    C_m+L/(T_2-T_1 ),& \text{if } T_1<T<T_2\\
    C_l,              &\text{if } T_2 \leq T
\end{cases}
\end{equation}
Where $T_1$ and $T_2$ mark the temperature range over which the melting occurs, and L, is the latent heat. For simplicity, $C_m$ is chosen to be ($C_s+C_l$)/2 here. This also demonstrated in figure \ref{fig:SPS_3} for YSZ suspension.

Surface tension coefficient also undergoes various changes during the evaporation of the liquid phase. As a wide-ranging set of information on surface tension versus concentration is not available for ceramic suspensions, in the current study, when liquid phase is present in droplets, surface tension is approximated to have the same value as that of the liquid phase. When the liquid phase is evaporated and the solid content has melted, the surface tension of the molten ceramic is used thereafter. In between these two limits, no breakup of the droplets is allowed and hence the surface tension values are not of interest.

\subsection{Effective viscosity}

\begin{figure}[H]
        \centering
                \includegraphics[width = 6in]{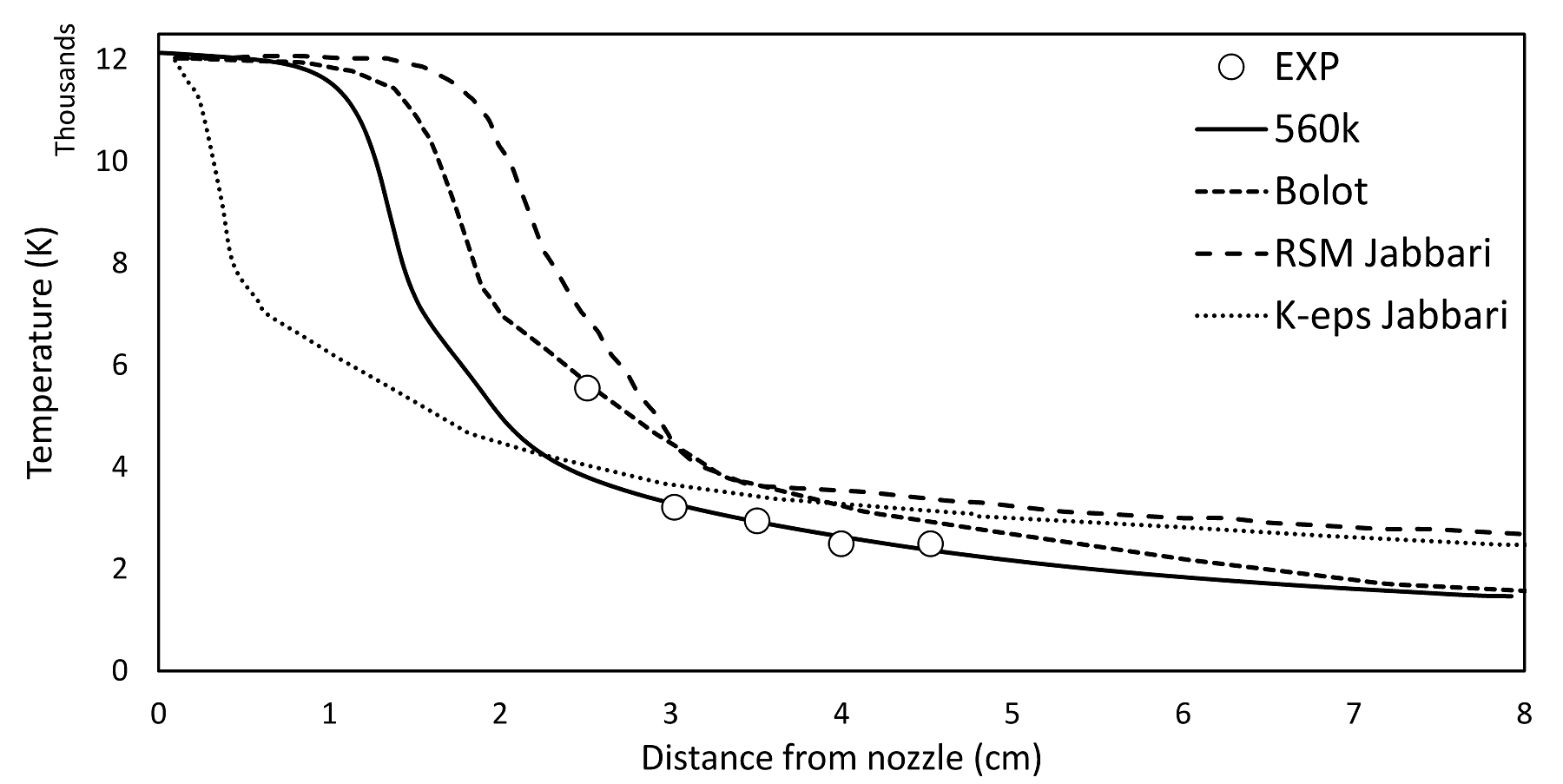}
\caption{Viscosity of YSZ-Water suspension under different concentrations of YSZ powder (from Arevalo-Quintero et al. \cite{Arevalo_11}). Dashed line is the trend line.}
\label{fig:SPS_4}
\end{figure}

With the evaporation of the liquid phase, the concentration of solid matter in droplets changes. Consequently, the viscosity of the mixture needs to be updated. Many studies have used analytical and empirical tools for evaluating suspension viscosities at different solid content concentrations. Einstein’s viscosity model is among the first that predicts viscosity of a dilute suspension of spherical droplets using \cite{Einstein_06}
\begin{equation}\label{eq:SPS_25}
\mu_r = 1 + 2.5 \phi
\end{equation}
with $\phi$ being the volume fraction of solid particles in suspension ($V_{solid}/V_{total}$). For higher concentrations, this equation was modified by Guth et al. \cite{Guth_36} to take interaction between solid particles into account in the form of
\begin{equation}\label{eq:SPS_26}
\mu_r = 1 + 2.5 \phi + 14.1 \phi^2
\end{equation}
where $\mu_r$ is the relative viscosity of the suspension to the pure fluid viscosity, shown as
\begin{equation}\label{eq:SPS_27}
\mu_r=\frac{\mu_{eff}}{\mu_{liq}}
\end{equation}
Using empirical data, Thomas enhanced this model to form \cite{Thomas_65}
\begin{equation}\label{eq:SPS_28}
\mu_r = 1+2.5 \phi + 10.05 \phi^2 + A \phi ^ {B \phi}
\end{equation}
where $A= 0.00273$ and $B=16.6$. For larger particles at higher concentrations and by taking effects of particle interactions into account, Toda el al. \cite{Toda_06} proposed using
\begin{equation}\label{eq:SPS_29}
\mu_r=\frac{1+0.5 k \phi - \phi }{(1-k \phi)^2 (1-\phi)}
\end{equation}
$k$ is a parameter obtained using empirical results and has been reported to be in the form of $k=1+0.6 \phi$ for spherical particles in water (~5$\mu$m radius). Models mentioned above do not take effects of maximum packing fraction, $\mu_m$, into account. $\phi_m$ is the largest possible volume fraction that can be achieved by adding solid particles to the suspension. Physical shape of particles along with their electric charge can affect the value of $\phi_m$. Since theoretically suspension viscosity should converge to infinity for solid concentrations close to $\phi_m$ \cite{Chong_71}, it is important to include this parameter into the model. Krieger et al. \cite{Krieger_59} have proposed inclusion of $\phi_m$ in the form of
\begin{equation}\label{eq:SPS_30}
\mu_r = {1 - \frac{\phi}{\phi_m}}^{- [\mu] \phi_m}
\end{equation}
Here $[\mu]$ is the intrinsic viscosity defined as the limiting value, which dominates suspension behaviour at low concentrations \cite{Shook_15} and is given by \cite{Schramm_96}
\begin{equation}\label{eq:SPS_31}
[\mu] = \lim_{\phi \to 0} \lim_{\dot{\gamma} \to 0} \left[ \frac{\mu_r-1}{\phi} \right]
\end{equation}
It is common to approximate $[\mu]$ to 2.5, which is derived by applying equation \ref{eq:SPS_31} to Einstein’s model in equation \ref{eq:SPS_25}. Dabak et al. \cite{Dabak_86} also proposed a model containing another empirically adjustable variable, $N$, given as
\begin{equation}\label{eq:SPS_32}
\mu_r=\left[1+\frac{[\mu]\phi}{N \left(1-\frac{\phi}{\phi_m} \right)} \right]^N
\end{equation}
$N$ is a flow related variable and is taken to be 2 for high shear rates \cite{Eilers_41}. More recently, Senapati et al. \cite{Senapati_10} has proposed an improved model that takes effects of particles size distribution along with shear rate into account
\begin{equation}\label{eq:SPS_33}
\mu_r=S \left[ 1+\frac{[\mu]}{\dot{\gamma}^n}  \frac{\phi}{\phi_m-\phi} \right]^N
\end{equation}
where $\dot{\gamma}$ is the shear rate, and n is the flow behavior index in the power law equation ($\tau=K\dot{\gamma}^n$). Parameter $S$ is calculated from $S=10C_Ud50$ where $C_U$ is the coefficient of uniformity calculated from $C_U=d60⁄d10$.
Predictions of models mentioned above show deficiencies when were tested here against available data for ceramic suspensions. A more recent model by Horri et al. \cite{Horri_11} however shows close predictions for ceramic suspensions. The relative viscosity of Horri, which is valid over a wide range of concentrations and shear rates, is in the form of
\begin{equation}\label{eq:SPS_34}
\mu_r=1+2.5\phi+K\phi \left( \frac{\phi}{\phi_m-\phi} \right)^2
\end{equation}
Parameter $K$  is determined using experimental data and varies for different shear rates. For 8 mol\% YSZ-water suspension using experimental results of Arevalo-Quintero et al. \cite{Arevalo_11} (as plotted in figure \ref{fig:SPS_4}, $\rho_s$=5.54g ⁄($cm^3$) ,  $\rho_l$=0.997 g ⁄($cm^3$) , $\mu_l$=0.89mPa⋅s), under a constant shear rate with $\phi_m$=21.76 , parameter $K$  can be estimated to be $K=5.660117$. As equation \ref{eq:SPS_34} produces good fits to experimental measurements of ceramic suspensions of interest here, it has been used in this study for prediction of ceramic viscosities. 

\subsection{Maximum packing fraction ($\phi_m$ )}

Most correlations used for viscosity approximation rely heavily on a good approximation of the maximum packing fraction. Two different methods of finding $\phi_m$  are reviewed here. The first method will be employed when experimental results for suspension viscosity at high concentrations is available. Where experimental data is missing, analytical models that are described below can be used instead.

\subsection{$\phi_m$ calculation from emperical data}

\begin{figure}[H]
        \centering
                \includegraphics[width = 6in]{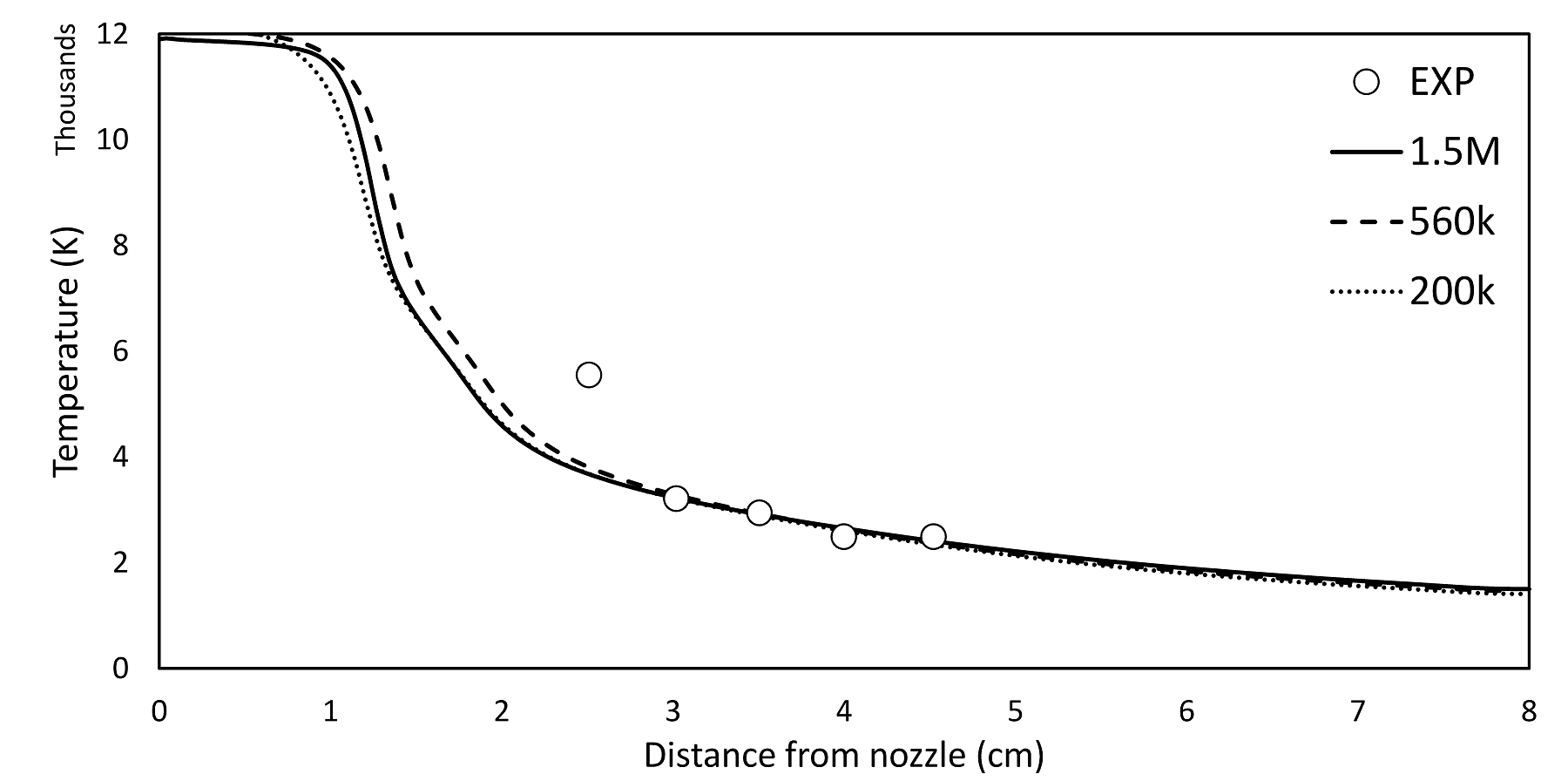}
\caption{$1/\mu_r$ for the YSZ-water suspension. Dashed line is the trend line.}
\label{fig:SPS_5}
\end{figure}

For the case where experimental data for suspension viscosity is available, the method explained by Senapati et al. \cite{Senapati_10} can be used to approximate $\phi _m$. Experimental results of Arevalo-Quintero et al. \cite{Arevalo_11} for viscosity of YSZ suspension in different solid contents (figure \ref{fig:SPS_4}) is used here as an example. The suspension used in their experiment was prepared using water and 8 mol\% YSZ powder with d10, d50, and d90 values close to 0.8$\mu m$ , 2.6$\mu m$, and 6.3$\mu m$, respectively.
The limiting value of $1/\mu_r$ when it goes to zero corresponds to $\phi_m$. Using values presented in figure \ref{fig:SPS_4} for viscosity of YSZ suspensions, $1/\mu_r$ can be calculated as shown in figure \ref{fig:SPS_5}. Using a trend line, or trial and error, the place for which $1/\mu_r$ has a very small value can be found here as $\phi_m$ = 21.76.

\begin{figure}[H]
        \centering
                \includegraphics[width = 6in]{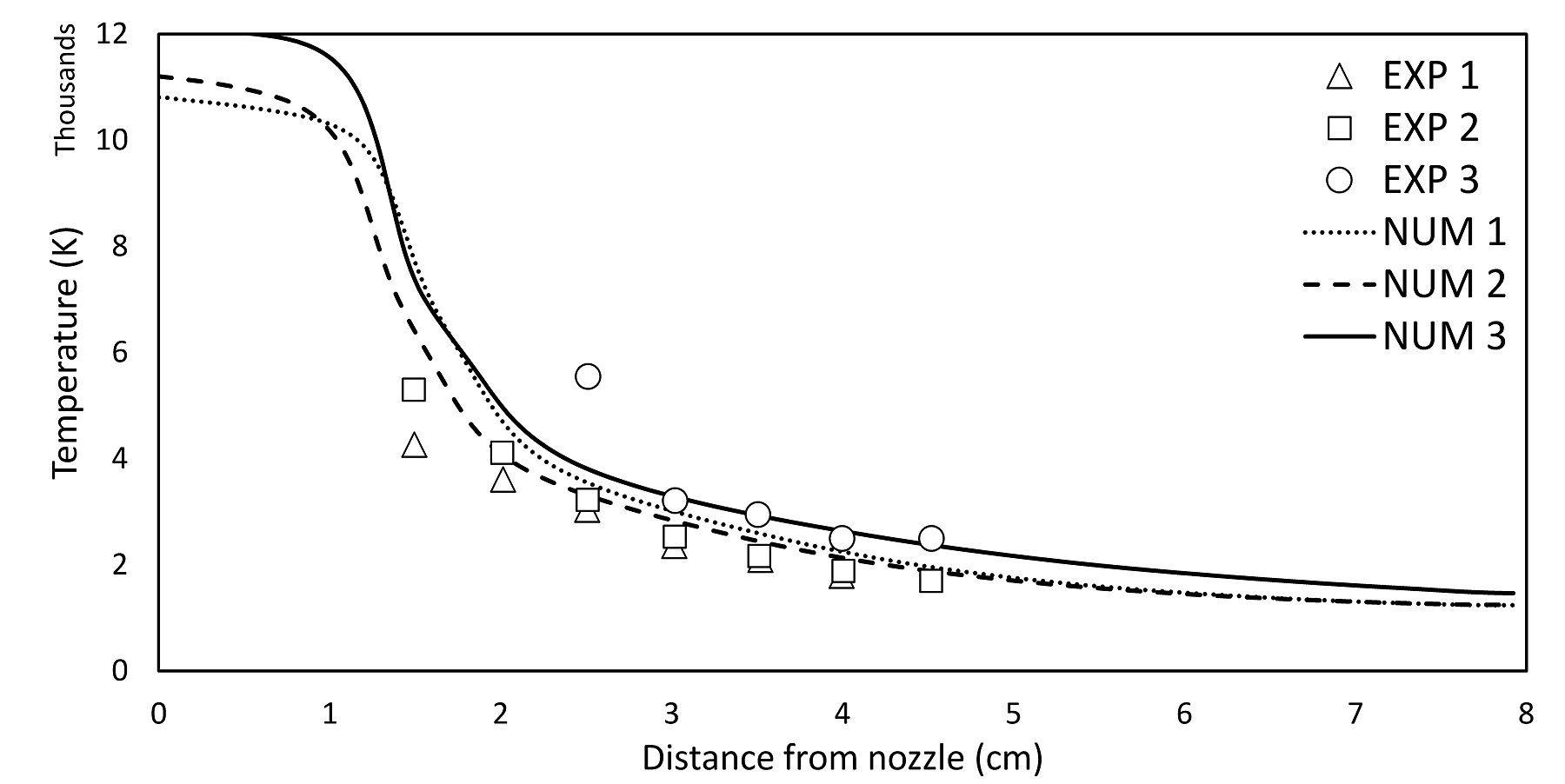}
\caption{$(1-\mu_r)^{-1/N}$ for the YSZ-water suspension. Dashed line is the trend line.}
\label{fig:SPS_6}
\end{figure}

Parameter N, appearing in equations \ref{eq:SPS_32} and \ref{eq:SPS_33}, is also a suspension dependant parameter and needs to be calculated from experimental data. As indicated by Senapati et al. \cite{Senapati_10}, this can be done using  $\left( 1-\mu_r\right)^{-1/N}-\phi$ curve. On this chart, which is shown in figure \ref{fig:SPS_6}, the value of $\phi$ corresponding to upper limit of $\left( 1-\mu_r\right)^{-1/N}=1$ should be identical to $\phi_m$=21.76 which was calculated earlier. By taking N = 1, $\phi$ = 21.76 leads to $\left( 1-\mu_r\right)^{-1/N}=1$, which is the same as $\phi_m$. It should be noted that the proper choice for high shear plasma flows is N = 2. However, the value of N = 1 produces a better match to the experimental results, suggesting the experiments were performed at lower shear rates. In the calculations presented here, the choice of N = 2 will not result in a significant change of $\phi$ value ($\phi \approx 24$).

\begin{figure}[H]
        \centering
                \includegraphics[width = 6in]{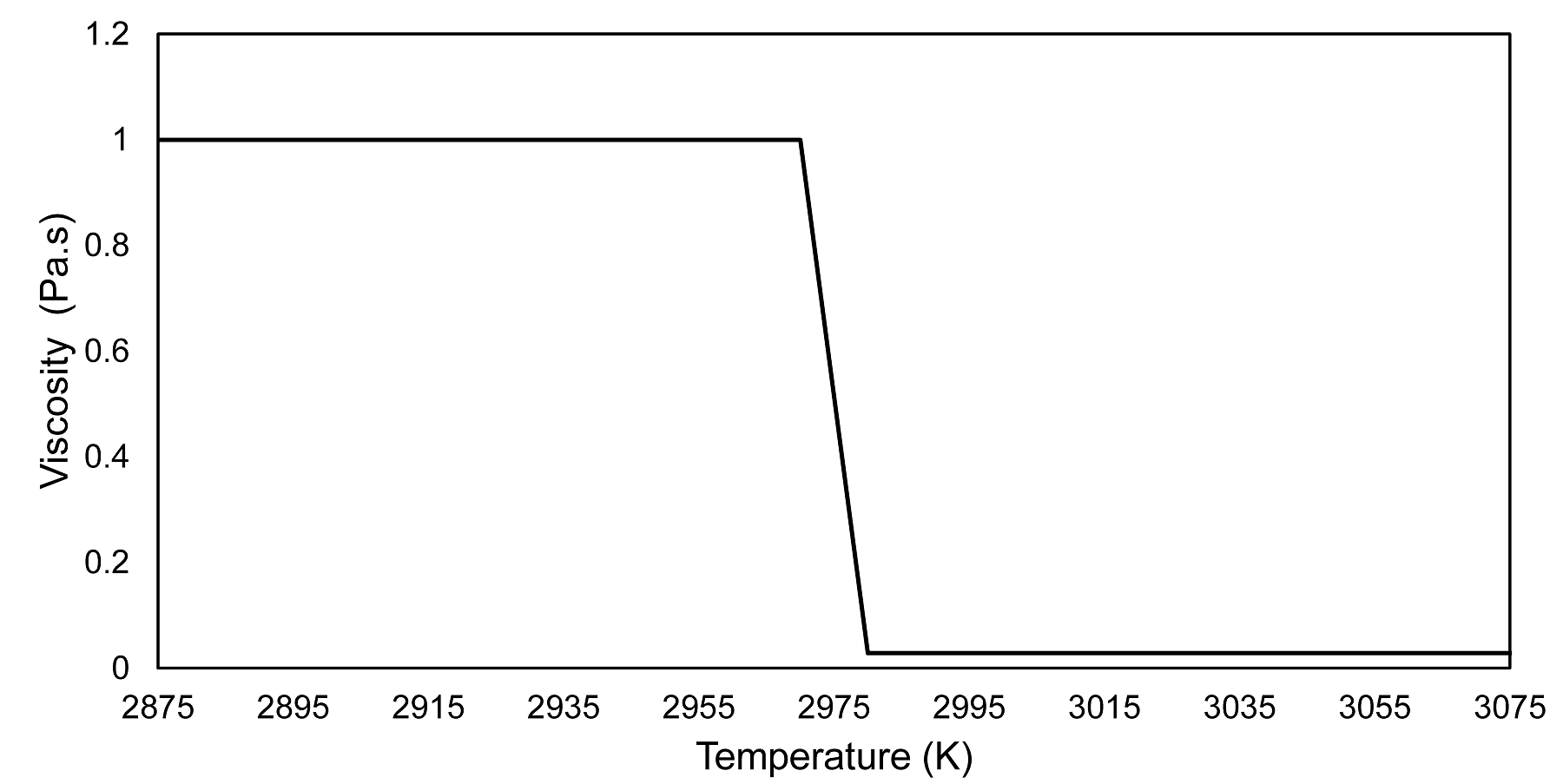}
\caption{Effect of turbulence model on temperature profile, experimental results are from Brossa et al. \cite{Brossa_88}  and numerical results from Bolot et al. \cite{Bolot_97} and Jabbari et al. \cite{Jabbari_14}.}
\label{fig:SPS_7}
\end{figure}

\begin{figure}[H]
        \centering
                \includegraphics[width = 6in]{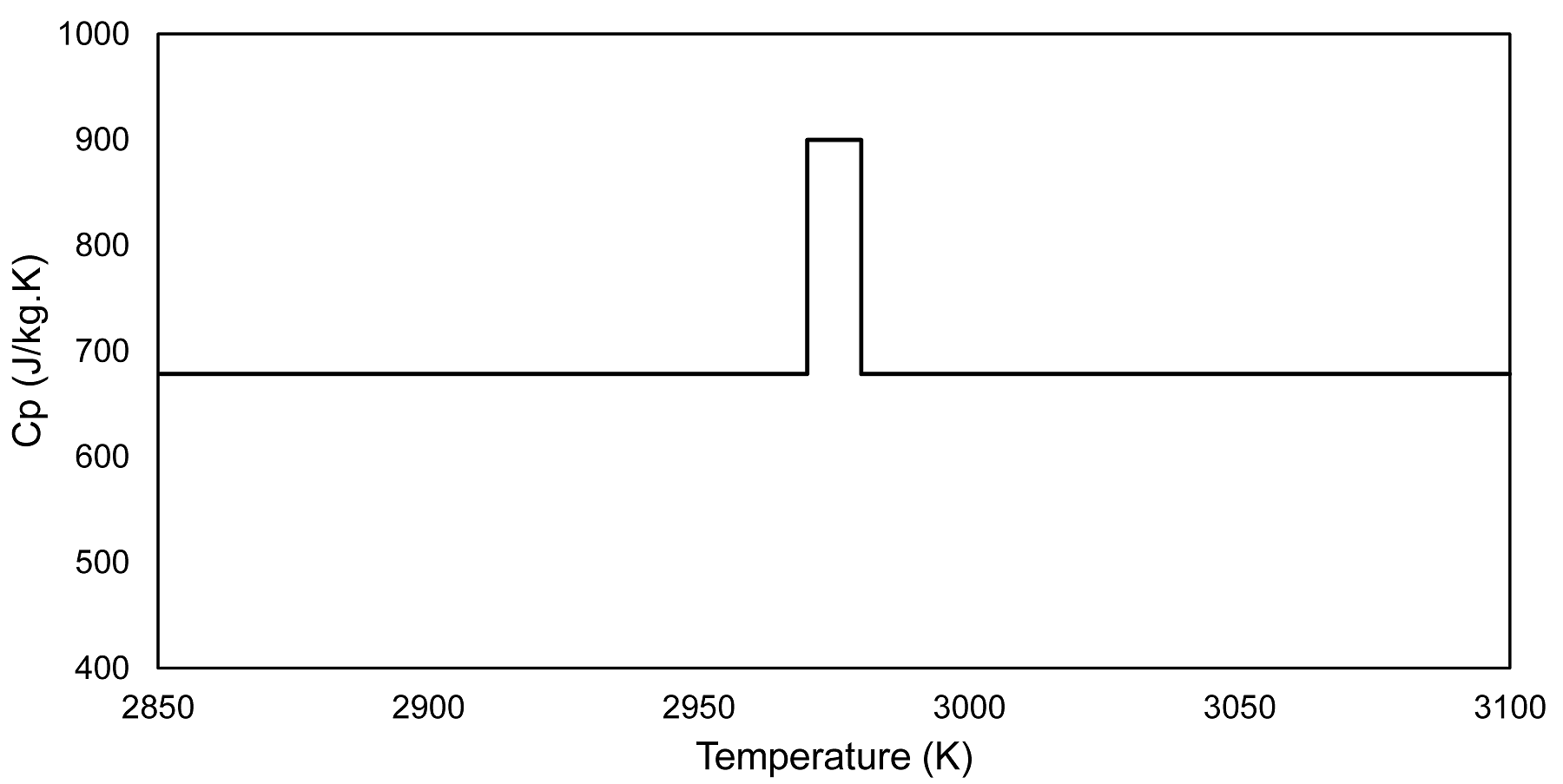}
\caption{Effect of mesh resolution on temperature profile, experimental results are from Brossa et al. \cite{Brossa_88}.}
\label{fig:SPS_8}
\end{figure}

\begin{figure}[H]
        \centering
                \includegraphics[width = 6in]{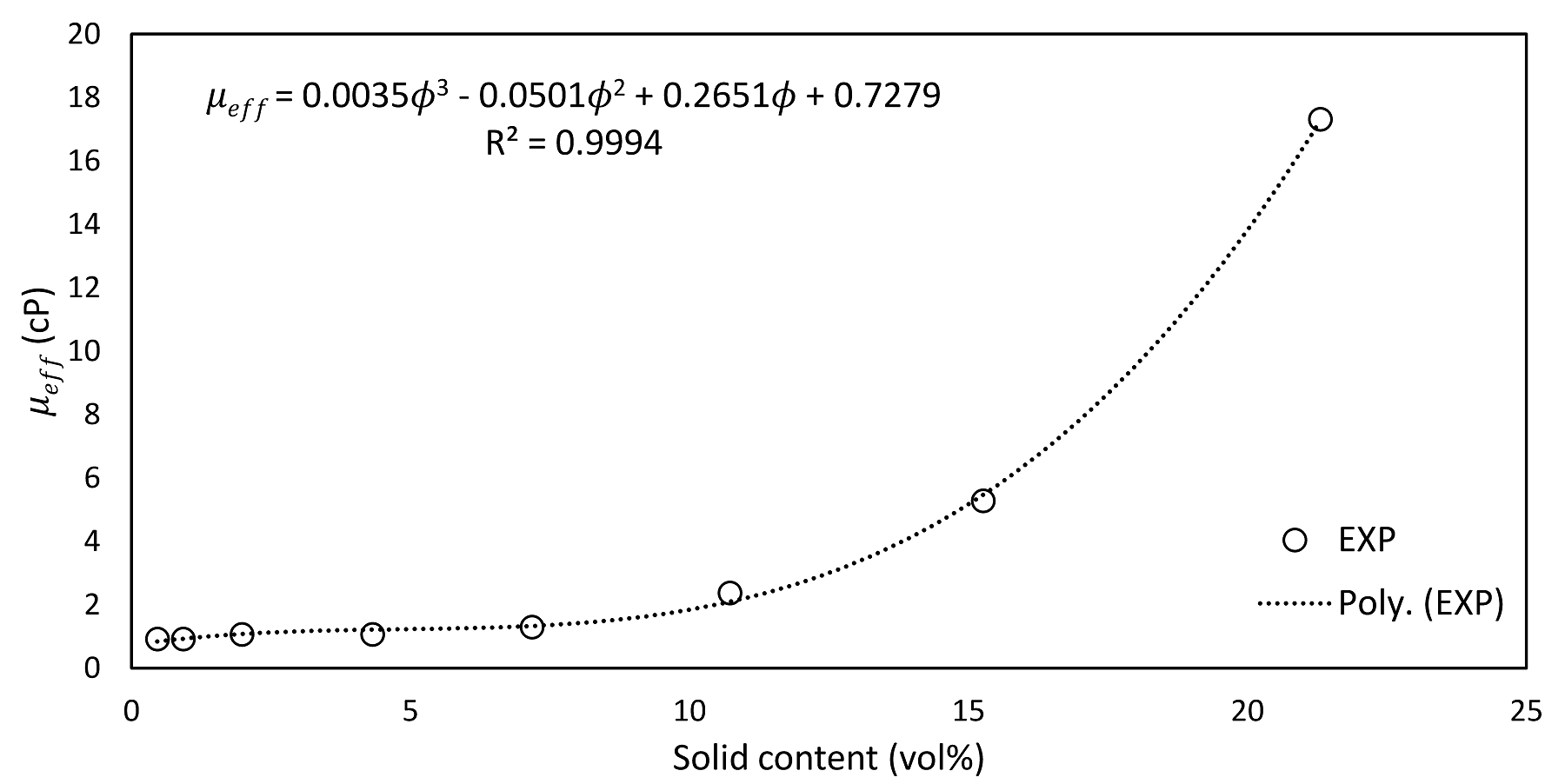}
\caption{Temperature profile for various torch conditions compared with experimental results of Brossa et al. \cite{Brossa_88}.}
\label{fig:SPS_9}
\end{figure}

\section{Results and Discussion}

\subsection{Torch Validation}

\begin{figure}[H]
        \centering
                \includegraphics[width = 6in]{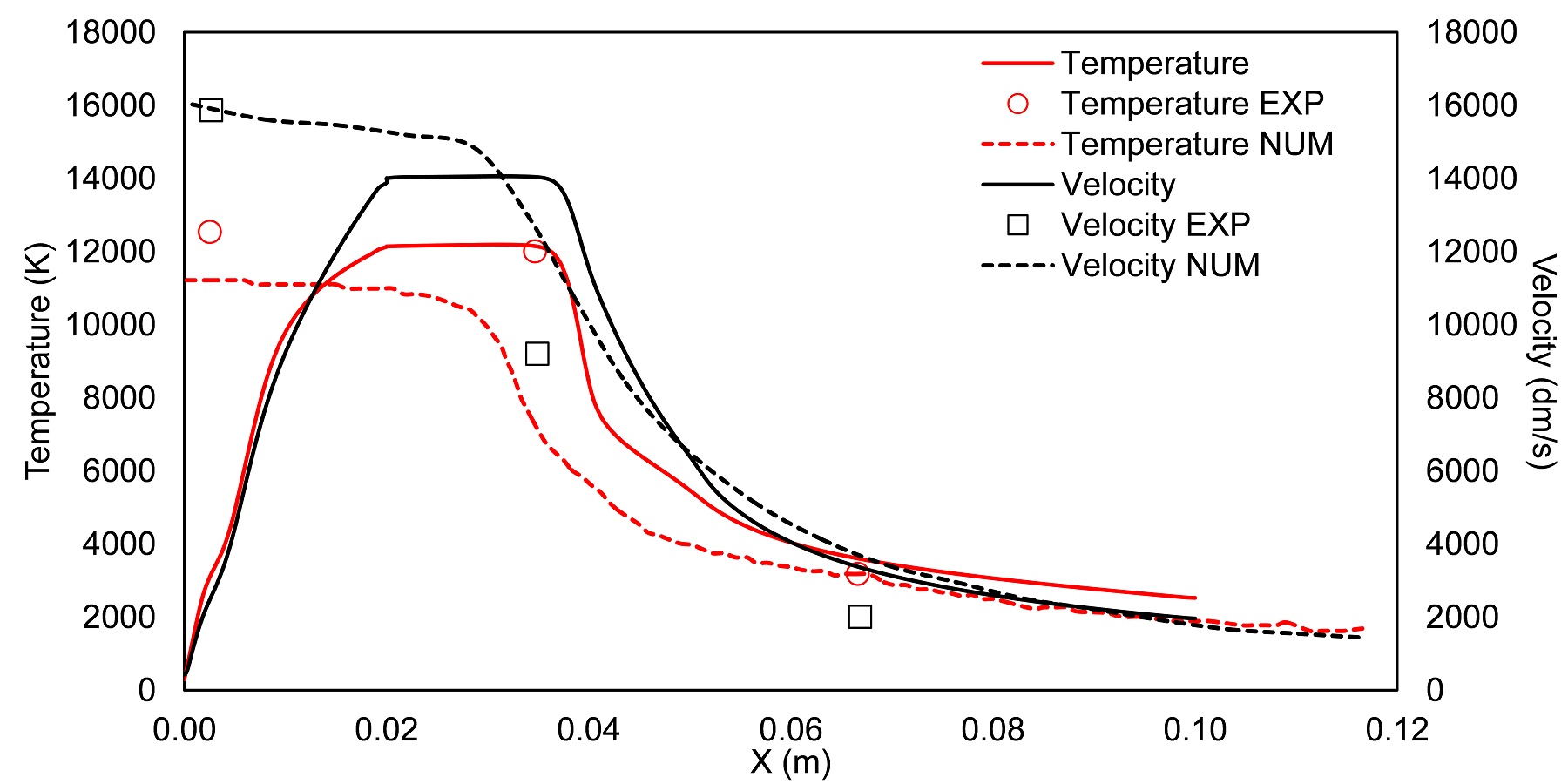}
\caption{Velocity and temperature profile on the torch centerline against experimental and numerical results of Meillot et al. \cite{Meillot_08}.}
\label{fig:SPS_10}
\end{figure}

Before continuing to particle injections, the torch model is benchmarked here against available experimental and numerical literature. For this test, a flow of Argon with 0.1 mole H2 is entered the torch at 35.4 slpm. Anode diameter is 6mm. Flow passing through anode is heated with a heat source corresponding a torch with 47\% thermal efficiency, 25.6 (V), and 600 (A). The domain is discretized using 560k cells. Fig. \ref{fig:SPS_7} shows results obtained here compared to other available test cases in literature. The k-ϵ results are a good match to the experimental results of Brossa et al. \cite{Brossa_88} and are also close to numerical predictions of Bolot et al. \cite{Bolot_97} as well as numerical RSM results of Jabbari et al. \cite{Jabbari_14}. 
The same test is repeated at different mesh resolutions to ensure the mesh independency. An extra fine mesh resolution (1.5M cells) and a coarser resolution (200k cells) are used for that matter. Results of these test cases are presented in Fig. \ref{fig:SPS_8}. As evident, the fine resolution of 560k cells produces close results to the extra fine resolution of 15.M. The rest of test cases in this paper are performed at this resolution.
The torch model has also been tested against experimental results of Brossa \cite{Brossa_88} for different torch operating conditions. Table \ref{tab:SPS_3} shows torch conditions for each of these test cases. Numerical results obtained here are plotted in Fig. \ref{fig:SPS_9}. The torch geometry used here is slightly different from Brossa’s experimental setup. The outer diameter of the Brossa’s torch is 7.88mm against a 6mm diameter here. Brossa’s torch also has a diverging nozzle which is not accounted for at this time. Numerical results here, similar to other studies, manage to capture the general temperature trend of the torch and are close to the experimental data. Closer accuracies can be achieved by using better implementations of the arc, which has been left for future studies. Another test case is also performed for comparison against results reported by Meillot et al. \cite{Meillot_08}. The operating conditions are 60slpm flow rate on a torch with arc current of 500A and a voltage of 65V. The value for thermal efficiency is 50\%. Results for this test case are shown in Fig. \ref{fig:SPS_10}.

\begin{table}[h]
\caption[Torch conditions]{Torch conditions}
\label{tab:SPS_3}
\begin{tabular*}{\textwidth}{c @{\hskip 0.3in \extracolsep{\fill}} ccccc}
 \toprule[1mm]
Case \# &	I (A)& V (V) & $\eta$ (\%) & $\dot{m}$(slpm)\\
 \hline
1	& 450       &27.2        &51.2       &47.2\\
2	& 450	&24.8	&47.1	&35.4\\
3	& 600	&25.6	&47.1	&35.4\\
 \toprule[1mm]
\end{tabular*}
\end{table}

\subsection{Effect of viscosity on breakup}

\begin{figure}[H]
        \centering
                \includegraphics[width = 6in]{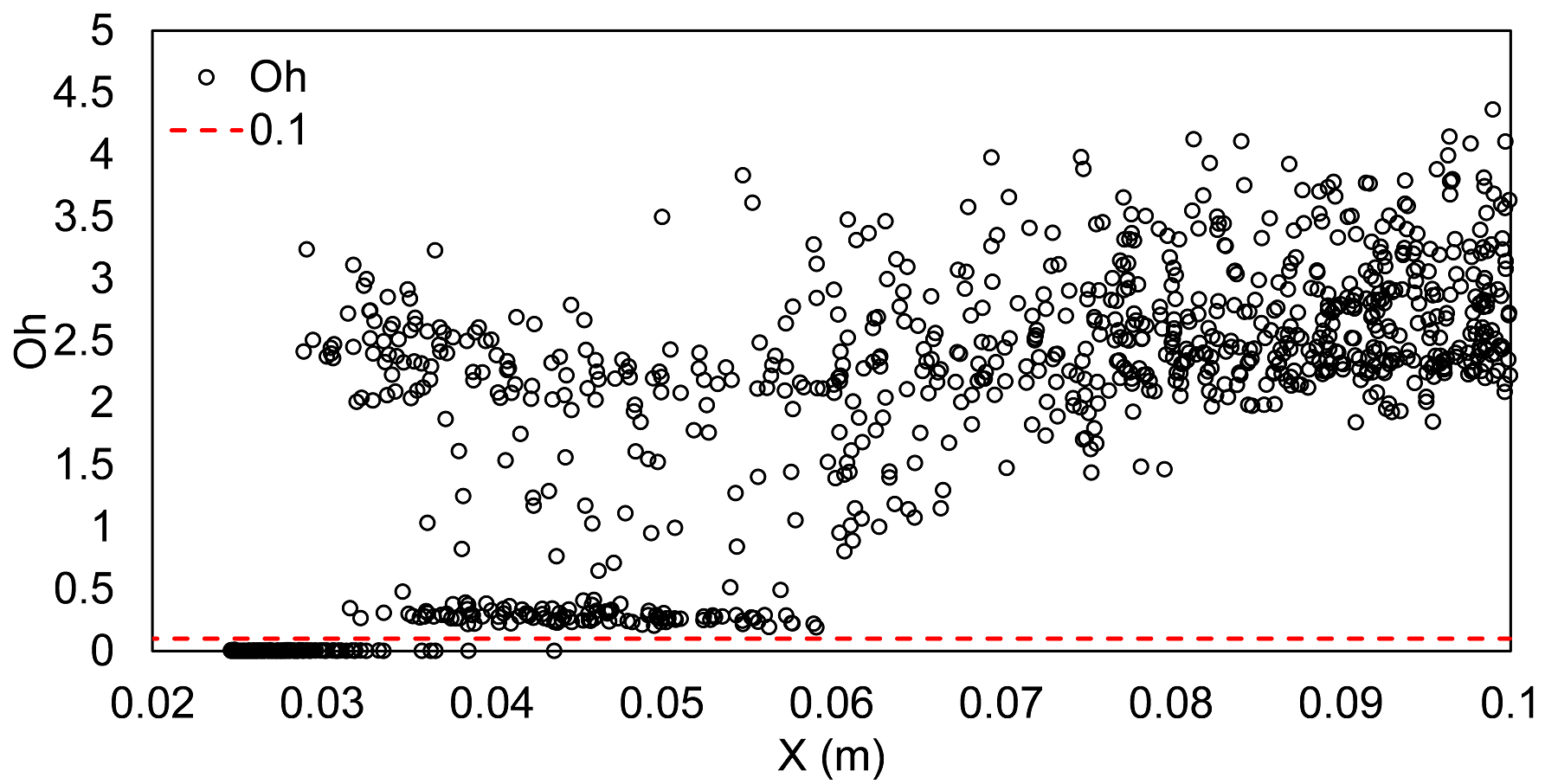}
\caption{Ohnesorge Number versus flight time for YSZ-water suspension particle injected into the domain.}
\label{fig:SPS_11}
\end{figure}

The importance of using a valid viscosity model to capture suspension droplets’ breakup can be emphasized by examining the test case described below. Torch operating conditions of case A (Table \ref{tab:SPS_5}) have been used in this test. Droplets injected are suspension of 10 wt\% YSZ in water and have a constant viscosity.
Ohnesorge number for injected droplets as they travel from the torch towards the substrate has been plotted in Fig. \ref{fig:SPS_11}. It is evident that many droplets have Oh Numbers larger than 0.1, which is the threshold from which the influence of viscosity becomes significant in the break up process \cite{Ashgriz_11}. With the assumption of constant viscosity, Oh numbers calculated here do not include effects of increase in viscosity as the solvent in the droplets evaporates. This neglected increase in viscosity will lead to even larger Oh numbers which will make the situation worse.

\subsection{Effect of viscosity model}

\begin{figure}[H]
        \centering
                \includegraphics[width = 6in]{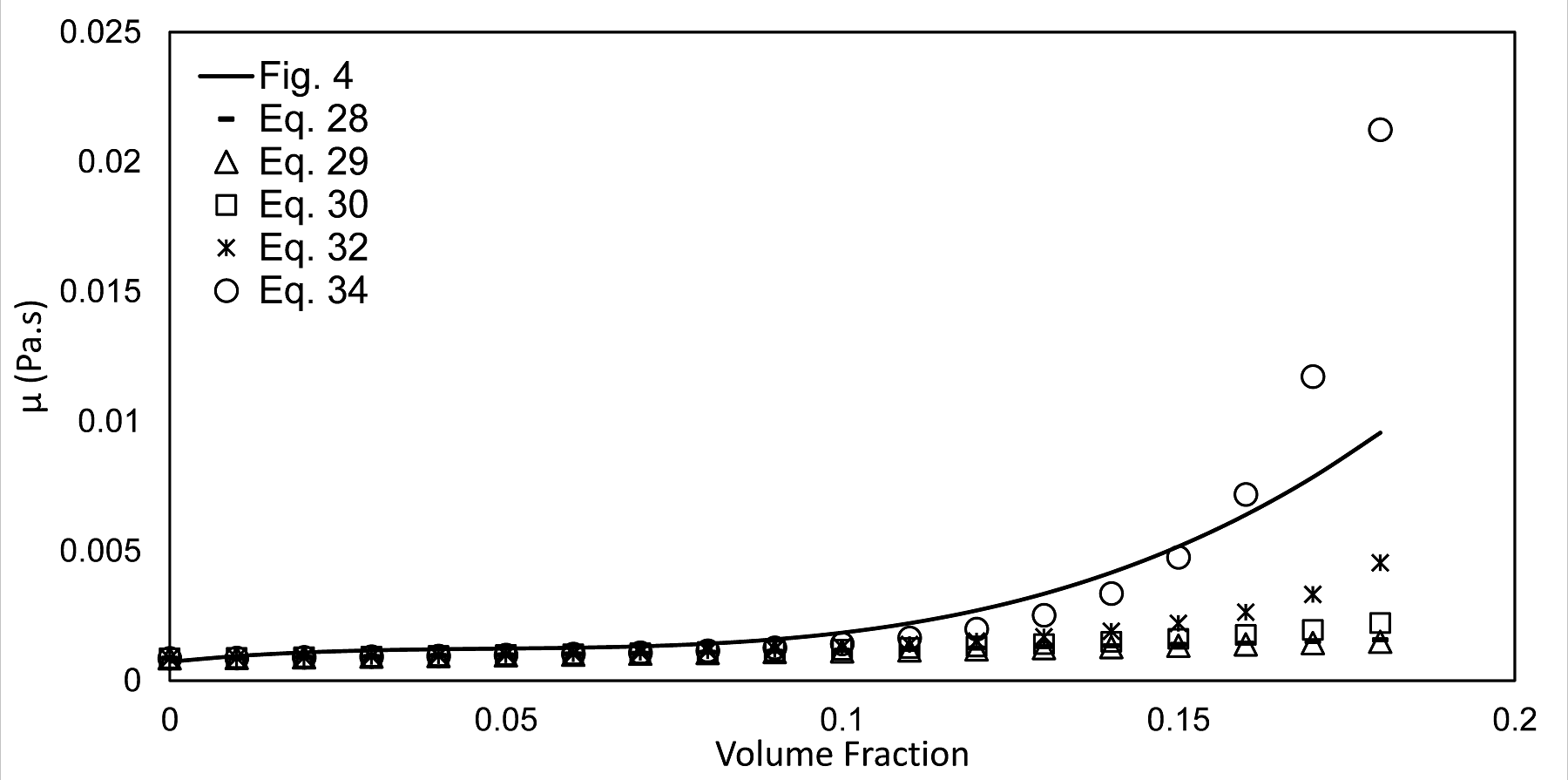}
\caption{Viscosity predictions for 8mol\% YSZ-water.}
\label{fig:SPS_12}
\end{figure}

As discussed before, a number of different models may be used for predicting effective viscosity based on instantaneous volume fraction of solids in suspension droplets. Since the experimental result for viscosity of 8 mol \% YSZ suspension in water at different concentrations is available (Fig. \ref{fig:SPS_4}), it has been used as a benchmark for the models described here. In one case, the experimental values for viscosity of this suspension have been used to update suspension viscosity at various solid concentrations. In this manner, at each time step in simulation, the volume fraction of solid content in each injected flying droplet is recalculated. This value is then used to update the viscosity for the suspension mixture of that particular droplet. This process is repeated for all injected droplets and during all iterations. Keeping all conditions identical, the same test case is repeated here, but instead, the viscosity of the suspension mixture is updated using equations \ref{eq:SPS_28}, \ref{eq:SPS_29}, \ref{eq:SPS_30}, \ref{eq:SPS_32}, and \ref{eq:SPS_34}. Equation \ref{eq:SPS_33} is not examined as suspension powder size distribution is neglected in the current model. Viscosity predictions from each equation has been plotted in Fig. \ref{fig:SPS_12} for different volume fraction values. Values close to $\phi_m$ have not been plotted, as predictions start growing to different large values, all being estimates for infinitely large viscosity. This figure suggests predictions by equation \ref{eq:SPS_34} are closer to experimental values of Fig. \ref{fig:SPS_4} compared to the rest of equations.

\begin{figure}[H]
        \centering
                \includegraphics[width = 6in]{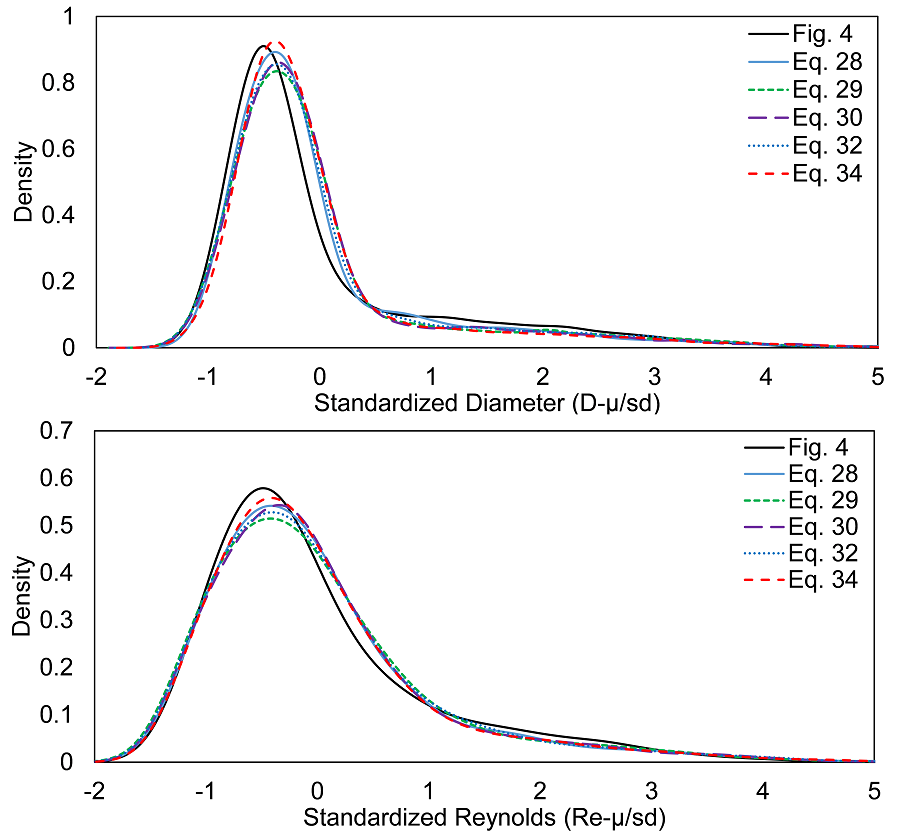}
\caption{Particle standardized diameter and Reynolds number density distribution for different viscosity models.}
\label{fig:SPS_13}
\end{figure}

Simulation results for these test cases have been summarized in Fig. \ref{fig:SPS_13}. Torch operating conditions here are that of case C in Table \ref{tab:SPS_5}. Particles traveling away from torch are here collected on a substrate located 8cm downstream of the nozzle exit. This figure shows the effect of viscosity model on the probability density distribution of particle diameter and Reynolds number upon impact on the substrate. Updating particle viscosities using equation \ref{eq:SPS_34} produces closer results to experimental viscosity data of Fig. \ref{fig:SPS_4}. Equations \ref{eq:SPS_28}, \ref{eq:SPS_29}, \ref{eq:SPS_30}, and \ref{eq:SPS_32} result in larger errors. Equation \ref{eq:SPS_29} is the least accurate.

\subsection{Effects of Injection Parameters}

\begin{figure}[H]
        \centering
  \begin{tikzpicture}[rotate=-90]

    \paficy{5}{10}{0.4}{0}
    \paficy{1}{2.5}{0.4}{0}
    \paficy{0.5}{2.5}{0.4}{0}

\draw [dashdotted] (0,0) -- (0,10);

\draw [dimen](0,5) -- (-2,5)node {$R$};
\draw [thin] (-1.05,2.5) -- (-2.1,2.5);
\draw [dimen] (-2,2.5) -- (-2,5) node {$H$};

\begin{scope}[shift={(-3.15,6.15)}]
    \paficy{0.1}{1.5}{0.4}{-135}
\end{scope}

\draw [thin] (-2.05,6.25) -- (-3.05,6.25);

\draw (-2.5,5.65) arc (-37:0:1);
\node[] at (-2.15,5.8)  {$\theta$};

\fill[ball color=black!20] (-1.8,4.8) circle (.1 cm);
\fill[ball color=black!20] (-1.5,4.5) circle (.1 cm);
\fill[ball color=black!20] (-1.2,4.2) circle (.1 cm);
\fill[ball color=black!20] (-0.9,3.9) circle (.1 cm);

\draw (.5,2.5) .. controls (0.5,4) and (0.3,6) .. (0,7.5);
\draw (-.5,2.5) .. controls (-0.5,4) and (-0.3,6).. (0,7.5);

\end{tikzpicture}
\caption{Schematic drawing of the parameters defining injection conditions.}
\label{fig:SPS_14}
\end{figure}
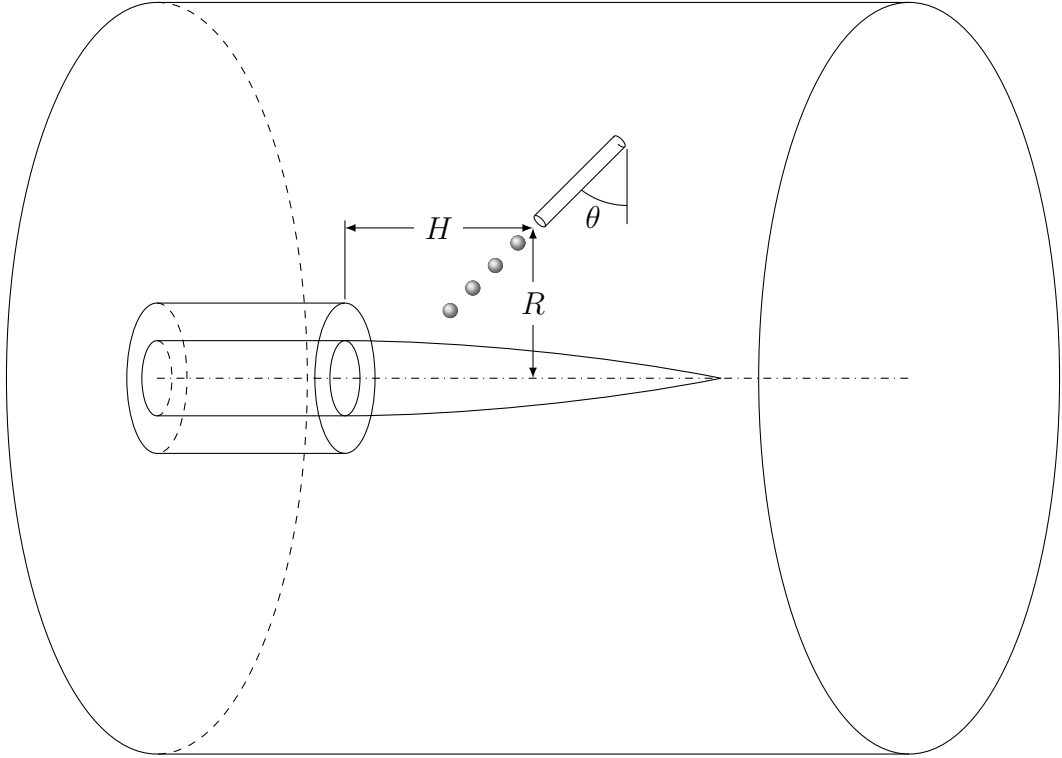

In the following section, effects of injection parameters on the particle trajectories are investigated. Droplets are injected with the assumption of being already atomized. For the moment, the size distribution of injected droplets is neglected and it is assumed that all droplets have the same diameter. Injection is controlled by four parameters: (i) Velocity magnitude of droplets at injection time ($v$), (ii) Injection angle ($θ$), (iii) Height ($H$), which is the injector’s distance from the torch exit, and (iv) Radial distance ($R$), which is the injector’s distance from the centreline of the torch. A schematic drawing of these parameters is shown in Fig. \ref{fig:SPS_14}.

After injection, droplets are tracked and undergo breakups, evaporation, or they may coalesce with one another. The liquid phase evaporates during flight and then solid phase starts melting. To save computational time, radiation heat losses are not considered for the test cases in current section. Molten particles are carried towards the substrate. In place of the substrate, 8cm downstream of the torch, a wall is placed to capture the particles impacting the substrate. The torch used for the calculations here has a nozzle diameter of 6mm, thermal efficiency of 47\%, and is operated at 25.6 ($V$) and 600 ($A$). For Ar-H2 gas mixture, according to \cite{Chen_83}, the accommodation coefficient (a) and γ in equation \ref{eq:SPS_13} are assumed to be 0.8 and 1.411, respectively.

\begin{figure}[H]
        \centering
                \includegraphics[width = 6in]{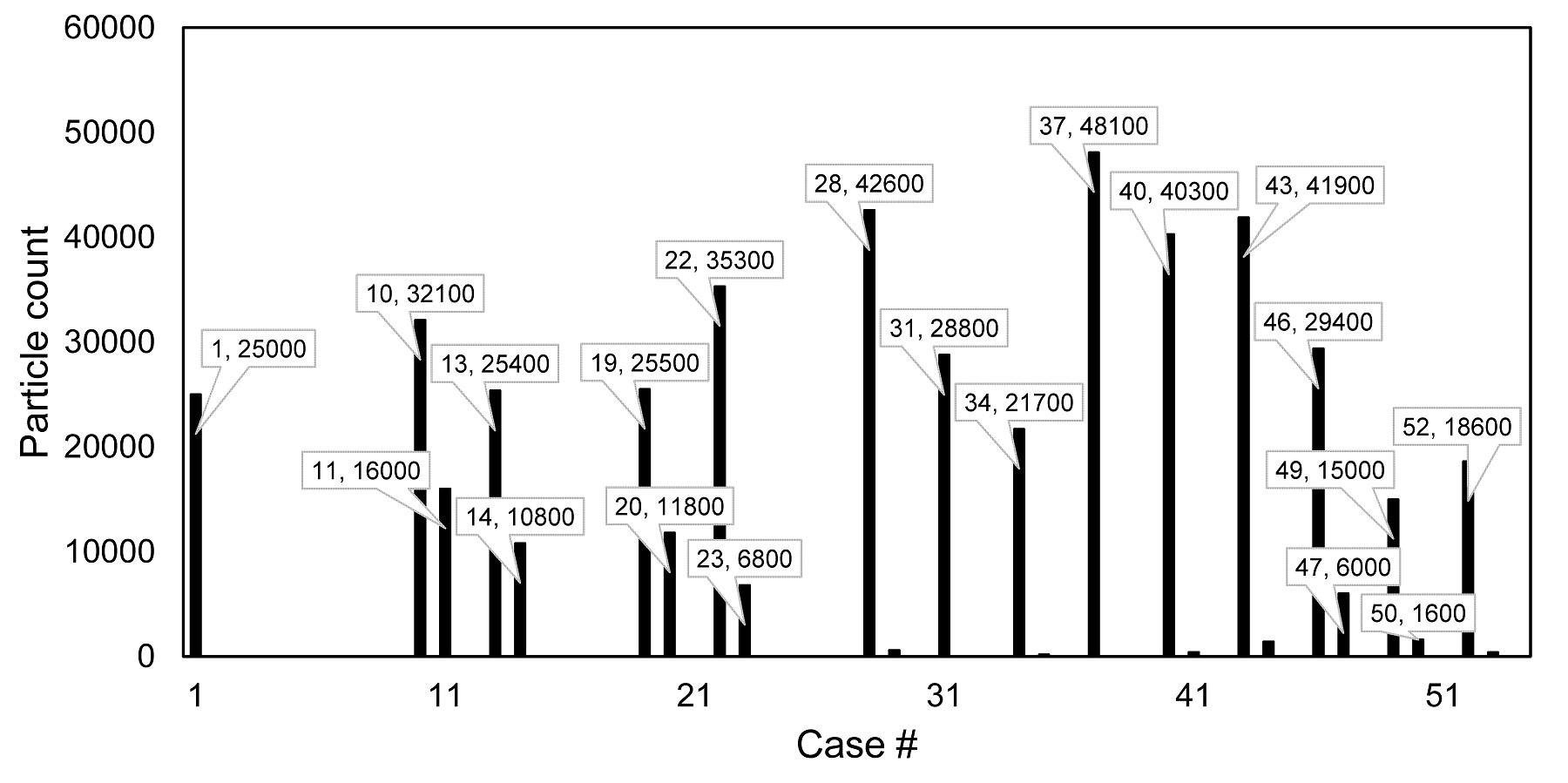}
\caption{Particle counts on substrate for each test case.}
\label{fig:SPS_15}
\end{figure}

For each test case in Table \ref{tab:SPS_4}, particles are captured on a substrate that is located 8cm from the nozzle exit. The spraying mass flow rate for all cases is kept the same. This means the time step for these test cases are different. The number of particles impacted the substrate during 1000 iterations is shown in Fig. \ref{fig:SPS_15}. Analysis of the particle trajectories is of interest here. We may examine how the injection parameters have led to particle penetration through the plasma jet. For many of the test cases, it is evident that injection parameters have caused the particles to miss the large substrate. This clearly shows the position of the torch and the mass flow rate of the suspension play important role in efficiency. Test cases 28-54, possess an overall larger particle count on the substrate compared to the rest of cases. This suggest that lowering the injection site closer to the torch centerline ($R=0$) will help improving particle counts on the substrate. This however may not be feasible for some torches and injectors. The results also show that increasing the injection velocity of particles, can work in two ways. For instance, keeping the rest of variables the same, increasing the velocity from 10 to 20m/s for test cases 13 and 14 has led to fewer particles reaching the substrate. The increase of velocity to 40 in case 15, has caused a complete penetration of the suspension particles and hence non have ended on the desired position. The same pattern can also be observed for cases 10-12, 19-21, 22-24, 46-18, and 49-51. For cases 4-9, 16-18, and 25-27 the injection is aimed too deep in the torch and hits inner and outer parts of the torch geometry and is scattered before successfully entering the main flow path. The largest number of particle count is related to case 37. This was predictable as in this case, the injection site is located on the centerline. Also the velocity of particles at injecting time is small compared to other cases, and hence gives the particles less chance for escaping the core flow. Test case 28 has also similar specification to case 37. The only difference is the injection site is moved from 1cm to 5mm from the nozzle exit. The moving of the injection site closer to the exit here has had a negative impact on the particle paths. This may indicate that the torch has recovered its symmetry better when the injection point was located at 1cm compared to 5mm. Moving the injection site downstream at 1.5cm, as in case 46, has also decreased the particle counts. This indicates the 1cm standoff is possibly closer to the ideal position of the injection site on the torch centreline.

\begin{table}[h]
\caption[Test case conditions]{Test case conditions}
\label{tab:SPS_4}
\centering
\resizebox{!}{.4\paperheight}{%
\begin{tabular*}{\textwidth}{c @{\extracolsep{\fill}} ccccc}
  \toprule[1mm]
Case \#&	R (m)	&H (m)&	$\theta$ (deg)	& v (m/s)\\
 \hline
1&0.025&0.005&10&10\\
2&0.025&0.005&10&20\\
3&0.025&0.005&10&40\\
4&0.025&0.005&20&10\\
5&0.025&0.005&20&20\\
6&0.025&0.005&20&40\\
7&0.025&0.005&40&10\\
8&0.025&0.005&40&20\\
9&0.025&0.005&40&40\\
10&0.025&0.01&10&10\\
11&0.025&0.01&10&20\\
12&0.025&0.01&10&40\\
13&0.025&0.01&20&10\\
14&0.025&0.01&20&20\\
15&0.025&0.01&20&40\\
16&0.025&0.01&40&10\\
17&0.025&0.01&40&20\\
18&0.025&0.01&40&40\\
19&0.025&0.015&10&10\\
20&0.025&0.015&10&20\\
21&0.025&0.015&10&40\\
22&0.025&0.015&20&10\\
23&0.025&0.015&20&20\\
24&0.025&0.015&20&40\\
25&0.025&0.015&40&10\\
26&0.025&0.015&40&20\\
27&0.025&0.015&40&40\\
28&0&0.005&0&10\\
29&0&0.005&0&20\\
30&0&0.005&0&40\\
31&0&0.005&15&10\\
32&0&0.005&15&20\\
33&0&0.005&15&40\\
34&0&0.005&20&10\\
35&0&0.005&20&20\\
36&0&0.005&20&40\\
37&0&0.01&0&10\\
38&0&0.01&0&20\\
39&0&0.01&0&40\\
40&0&0.01&15&10\\
41&0&0.01&15&20\\
42&0&0.01&15&40\\
43&0&0.01&20&10\\
44&0&0.01&20&20\\
45&0&0.01&20&40\\
46&0&0.015&0&10\\
47&0&0.015&0&20\\
48&0&0.015&0&40\\
49&0&0.015&15&10\\
50&0&0.015&15&20\\
51&0&0.015&15&40\\
52&0&0.015&20&10\\
53&0&0.015&20&20\\
54&0&0.015&20&40\\
  \toprule[1mm]
\end{tabular*}
}
\end{table}

\subsection{Effect of Torch Operating Conditions}

Plasma torch operating conditions play a key role in the effectiveness of the SPS process. Table \ref{tab:SPS_5} shows the range of operating conditions for the torch considered in this work, which include the Ar-H2 mass flow rate and arc voltage variations. For case A, B, and C, only the gas flow rate is varied.  Torch power is also varied using the voltage value. For cases A, D, and E, the change in voltage generates power variations of 7,235W, 14,470W, and 21,704W, respectively. The mass Flow rate for these three cases is kept identical (35.4slpm). In all test cases, the right boundary of the domain at 8cm is a large substrate.

\begin{table}[h]
\caption[Effect of torch parameters]{Effect of torch parameters}
\label{tab:SPS_5}
\begin{tabular*}{\textwidth}{c @{\extracolsep{\fill}} ccccc}
 \toprule[1mm]
Case \# &	I (A)& V (V) & $\eta$ (\%)  & P (W) & $\dot{m}$(slpm)\\
 \hline
A&600&25.6&47.1&7,235&35.4\\
B&600&25.6&47.1&7,235&70.0\\
C&600&25.6&47.1&7,235&140.0\\
D&600&51.2&47.1&14,470&35.4\\
E&600&76.8&47.1&21,704&35.4\\
  \toprule[1mm]
\end{tabular*}
\end{table}

In all these cases, the injected suspension is 10 wt\% YSZ-water. Droplets are injected with a uniform initial diameter of 150$\mu m$. The injector properties for these test cases are $R$=0.005m, $H$=0.005m, $\theta=10^{\circ}$, and $v$=10m/s. Injection flow rate is kept constant at 1.92E-4kg/s. The suspension viscosity is corrected utilizing Fig. \ref{fig:SPS_4}. Evaporation of water, breakup and coalescence of liquid suspension and molten drops, and radiation heat losses are all taken into account.
Two particular aspects affecting the droplets/particles are of interest. First is the conditions during its flight, e.g., shear forces exerted on the droplets at each particular point of their flight path. The second area of interest is particle conditions upon impact on the substrate. Conditions such as particle temperature, diameter, and velocity play an important role in defining the final finish of the coating. In the following sections, for test cases A-E, initially droplet conditions during flight are examined. Afterwards, particle properties upon impact on the substrate are investigated.

\subsection{In-flight conditions}

\begin{figure}[H]
        \centering
                \includegraphics[width = 6.5in]{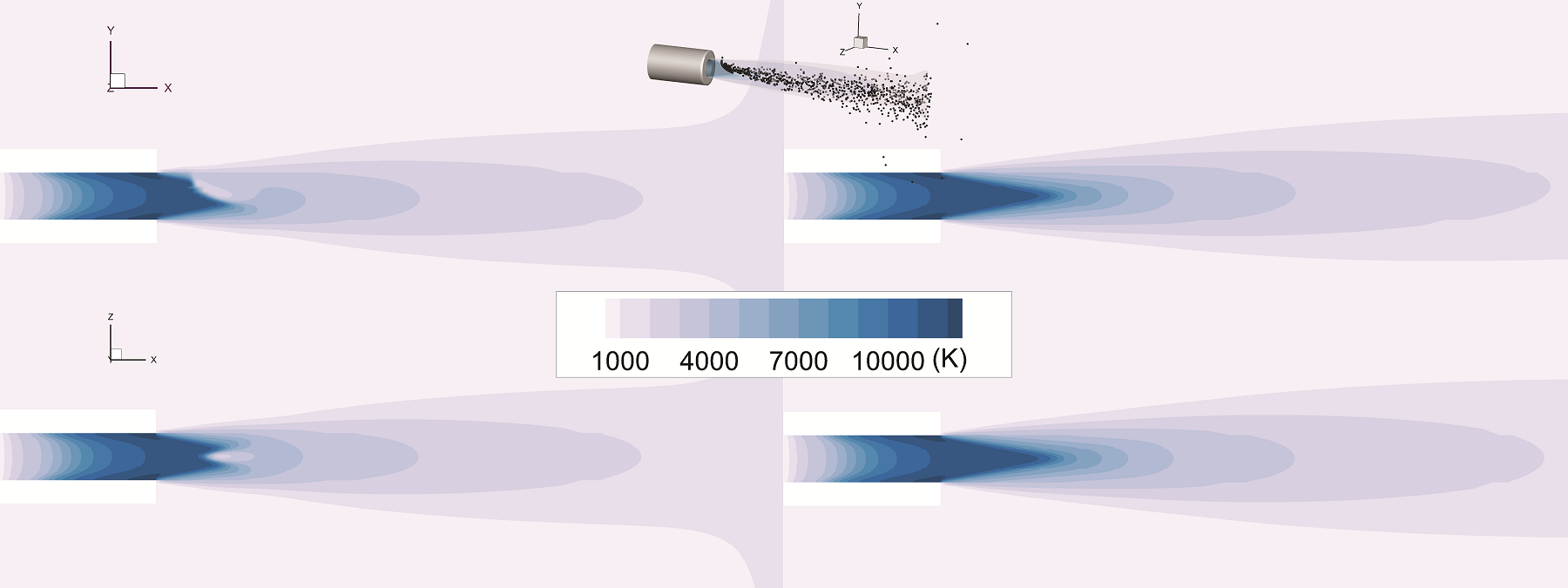}
\caption{Axial cross sections for the torch of case A.}
\label{fig:SPS_16}
\end{figure}

The injection of liquid suspension alters the flow pattern of the plasma gas. The momentum carried by the liquid jet traveling towards the torch divides the flow into two steams for a short distance. Temperature contours on axial cross sections of the torch from two perpendicular angles have been shown in Fig. \ref{fig:SPS_16} for case A. Images on the left show the torch with injection and substrate, while images on the right show the torch without injection and substrate. The liquid jet alters the torch flow pattern by slightly shifting the position of the maximum velocity at the nozzle exit. Also the liquid injected is typically at room temperature. This has a cooling effect on the plasma flow. More importantly, evaporation of the liquid content in the suspension is the major cause of cooling of the plasma.

\begin{figure}[H]
        \centering
                \includegraphics[width = 6in]{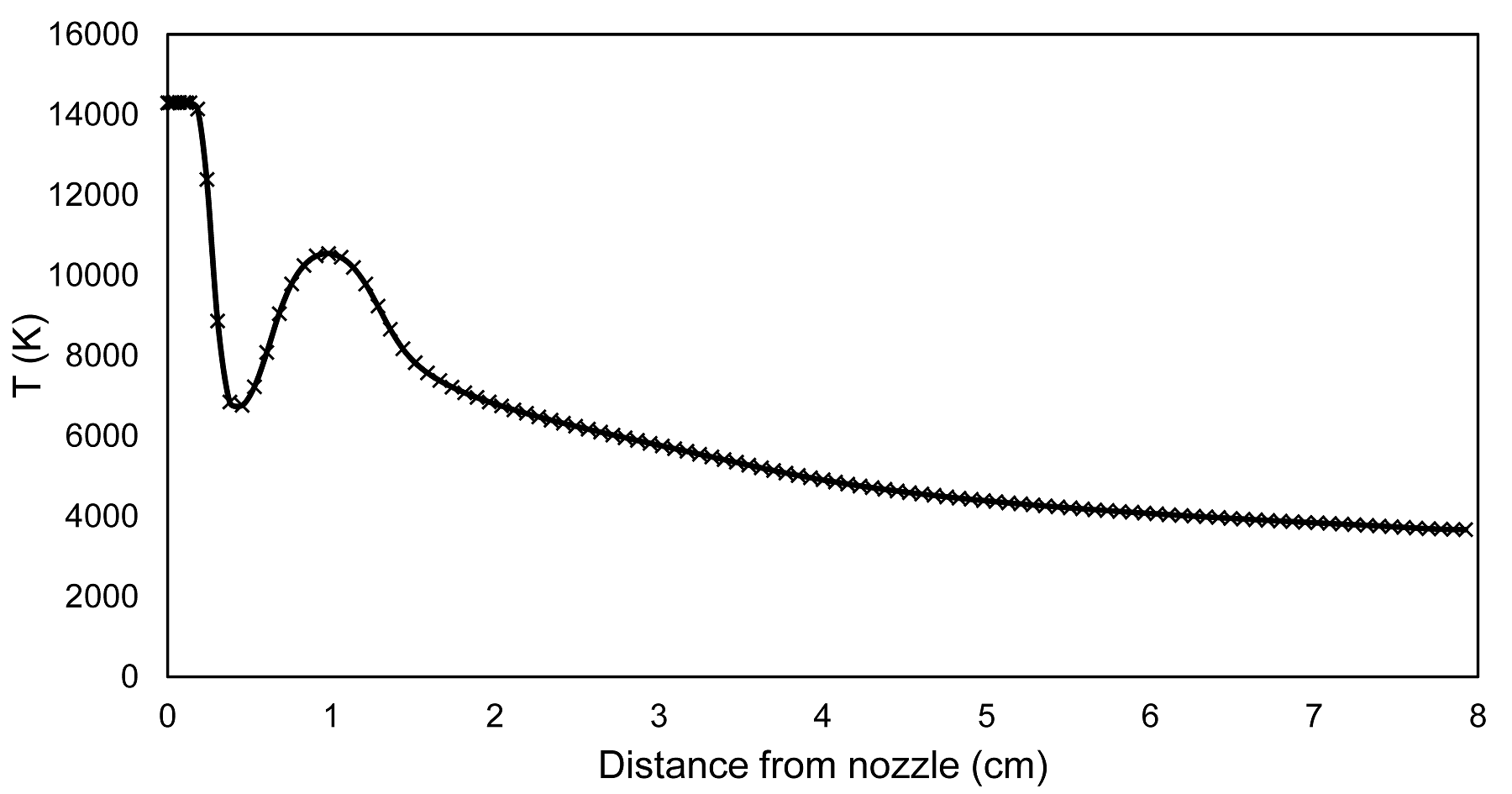}
\caption{Temperature profile of the plasma gas flow on the centerline of the torch (case E, Table \ref{tab:SPS_3}).}
\label{fig:SPS_17}
\end{figure}

\begin{figure}[H]
        \centering
                \includegraphics[width = 6in]{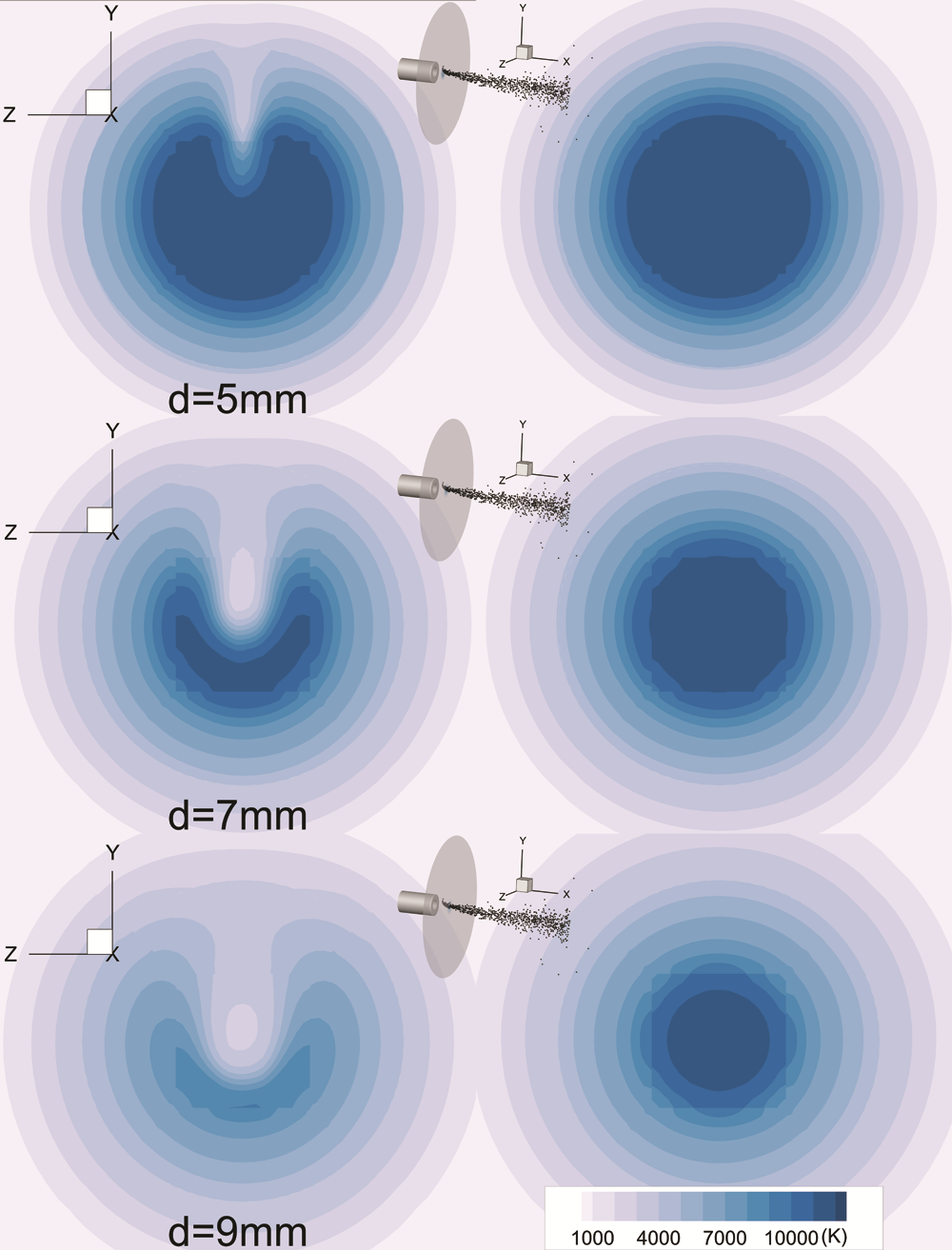}
\caption{Temperature profile for case A, plotted at cross sections perpendicular to the flow direction at distances of 3, 5, and 7mm from the nozzle exit (from top to bottom).}
\label{fig:SPS_18}
\end{figure}

\begin{figure}[H]
        \centering
                \includegraphics[width = 4in]{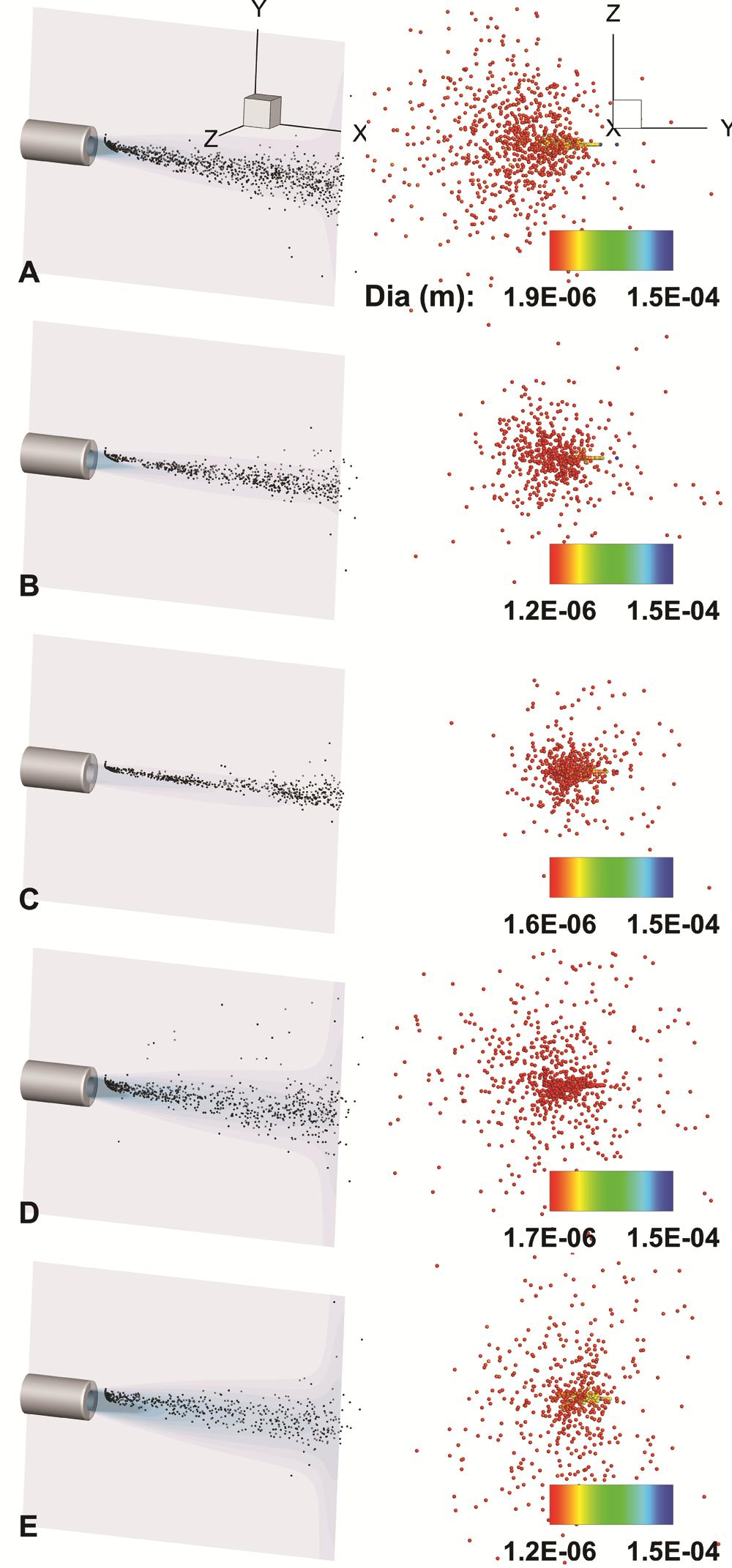}
\caption{Snapshot of particle positions during flight for Cases A-E. Images on the right show particles during their flight from substrate’s point of view. Particles are colored based on their diameter (m).}
\label{fig:SPS_19}
\end{figure}

As another example, temperature profile of the plasma flow for case E has been plotted in Fig. \ref{fig:SPS_17} on the centerline of the torch. This chart clearly shows the cooling effects of suspension on plasma. It can be seen that the flow near the injection penetration site has a sudden drop in temperature, which starts recovering after 5mm downstream where the flow symmetry is being restored. Fig. \ref{fig:SPS_18} shows the temperature field for case A at three cross sections, with distances of 5, 7, and 9mm from the nozzle exit. Images on the left are with injection, while images on the right are without injection. The injector here is aimed close to the center point of the nozzle exit. It is evident that due to existence of the liquid stream, the temperature profile is no longer symmetric at each cross section. Comparison of the temperature profiles shows the important cooling effect. This cooling is not significant before 5mm. However, as the liquid phase evaporates, the cooling becomes more pronounced, as can be seen in the cross section at 9mm. Similar to findings of Fazilleau et al. \cite{Fazilleau_06}, the torch here recovers its symmetry flow pattern after 1-2cm downstream of the nozzle exit. The maximum temperature of the flow however is lower after the symmetry is restored compared to when there is no injection. Fig. \ref{fig:SPS_19} shows a snapshot of particle positions during their flight for cases A-E.

\begin{figure}[H]
        \centering
                \includegraphics[width = 6.5in]{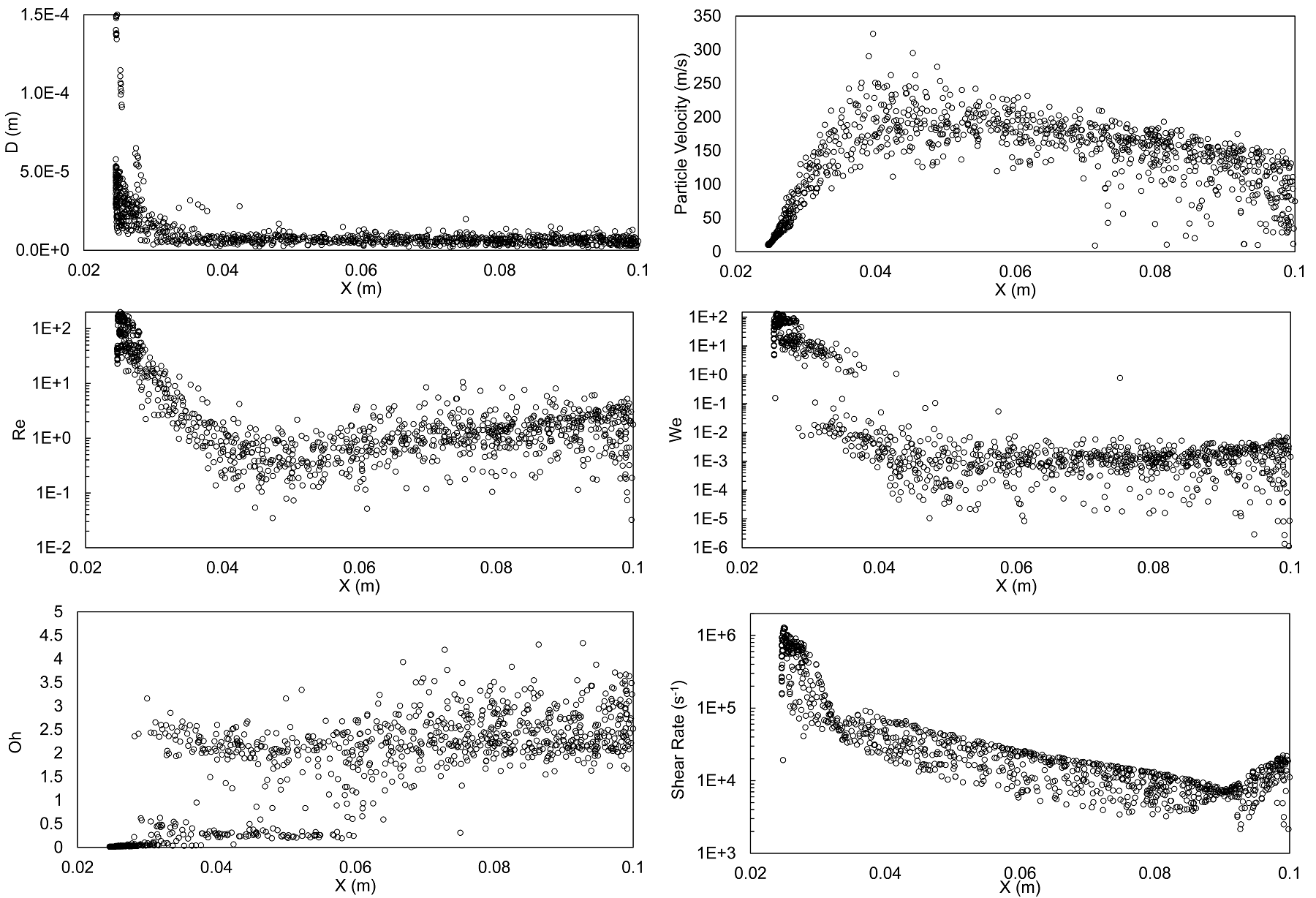}
\caption{Diameter, velocity, Reynolds number, Weber number, Ohnesorge number, and shear rates for particles during flight. Shown here for case A. The torch nozzle exit is located at X=0.02m.}
\label{fig:SPS_20}
\end{figure}

To study the effect of plasma-particle interactions more closely, other properties of particles for case A have been plotted in Fig. \ref{fig:SPS_20}. These properties are the diameter of each particle (D), mass fraction of water inside each particle that has not been evaporated yet, particle Reynolds number defined as $Re=\rho_g v_{rel} d_p / \mu_g$, particle Weber number defined as $We=\rho_g v_{rel}^2 d_p / \sigma_p$, particle Ohnesorge number defined as $Oh=\mu_p / \sqrt{\rho_p \sigma_p d_p }$, and shear/strain rate at the cell where the particle is located. The x-axis here indicates the distance each particle has travelled from the nozzle exit.

\begin{figure}[H]
        \centering
                \includegraphics[width = 6.5in]{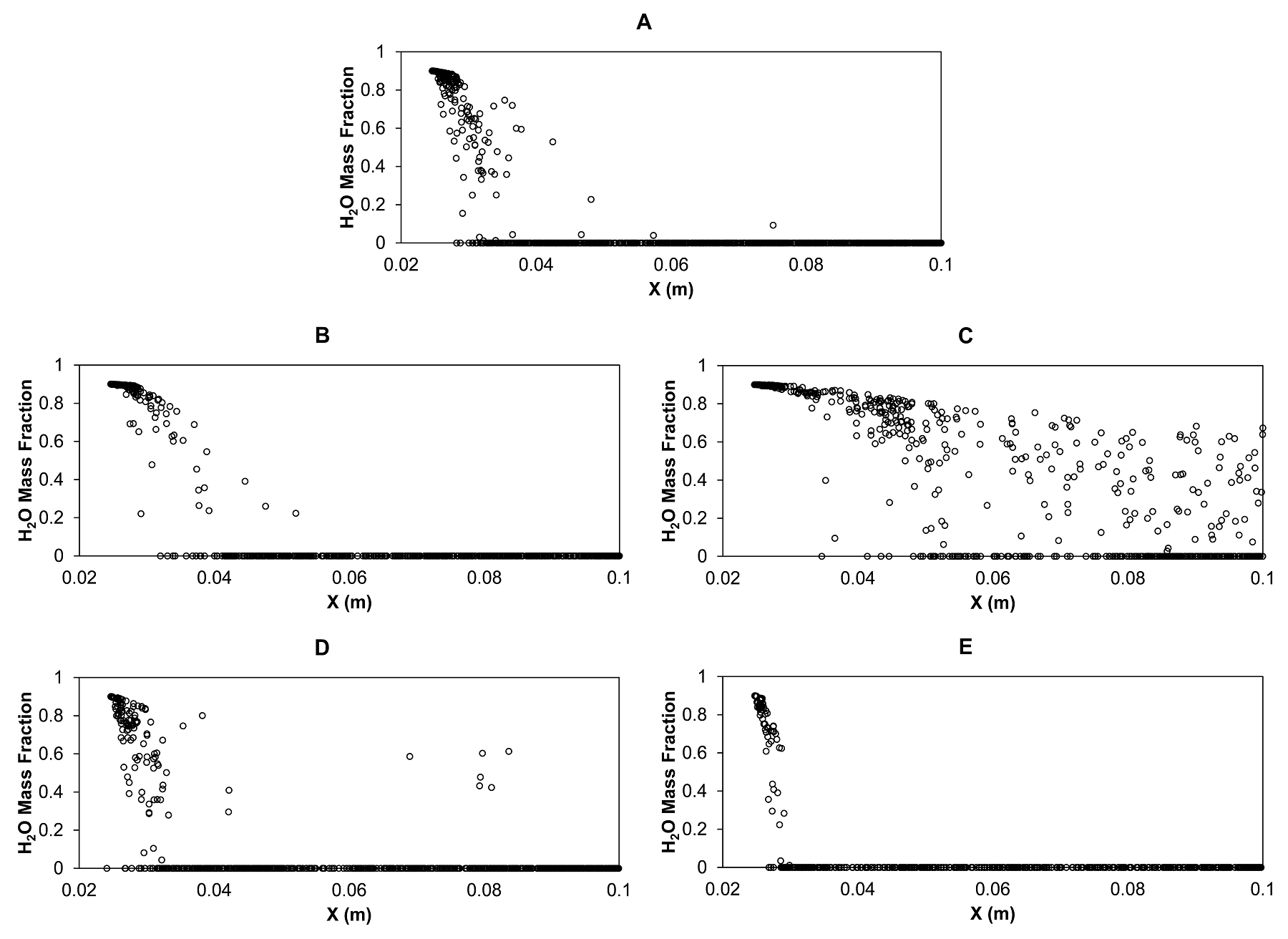}
\caption{Water mass fraction inside particles during flight for cases A-E.}
\label{fig:SPS_21}
\end{figure}

The Reynold and Weber diagrams of the droplets exhibits a huge jump near the nozzle exit and converges fast to smaller values as they move away from the nozzle exit. The ratio of these two has been plotted as the Ohnesorge number. The values of the Oh become large, especially after the liquid content is evaporated. This increase in Oh should be handled with caution as the apparent increase in viscosity, imposed here to numerically mimic the presence of solid YSZ particles, might be contributing to it. Hence, the values of these non-dimensional numbers must be studied next to particle temperatures. For case A, particles have temperature readings larger than melting point from approximately 15 to 33mm. Many particles are also at their melting temperature and might be in the mushy region from 18 to 51mm. Over this range, many particles have Reynolds numbers below 3. Small groups also have values close to 30, 50, and 150. The Weber numbers are all below 7 at this range, with many having values below 0.01. These values result in two Ohesorge numbers close to 0.3 and 3. Shear rate trends can also be seen in Fig. \ref{fig:SPS_20}. These trends are close for all test cases. The maximum is similar for test cases A and C. Cases B and E have slightly higher rates, while the largest is seen for case D. In these figures, small number of droplets can be observed having very low shear rates. These are the droplets that have either penetrated through the plasma jet flow at the injection point, or have escaped the flow and are travelling far from the centerline. If the injection parameters are not chosen correctly, the liquid suspension can penetrate pass the plasma jet and many of the droplets will end up escaping nearly untouched by not going through evaporation and breakups.
The liquid mass fractions inside droplets during flight is shown in Fig. \ref{fig:SPS_21} for cases A to E. The torch nozzle exit is located at X=0.02m here. Fazilleau et al. \cite{Fazilleau_06}  have reported that at 10-15mm downstream of the nozzle exit, all the liquid content in the suspension will be already evaporated. For test cases here, the liquid content in most of the injected droplets is evaporated nearly at that distance. For case C however, the evaporation of solvent takes longer. Compared to cases A and B, case C has the same amount of power, but also a higher mass flow rate. Hence the particles injected into this flow will have less time to absorb enough energy to finish their phase change cycle. Best performance is here is seen for case E, where all particles lose their liquid solvent and at the fastest rate.

\begin{figure}[H]
        \centering
                \includegraphics[width = 5.5in]{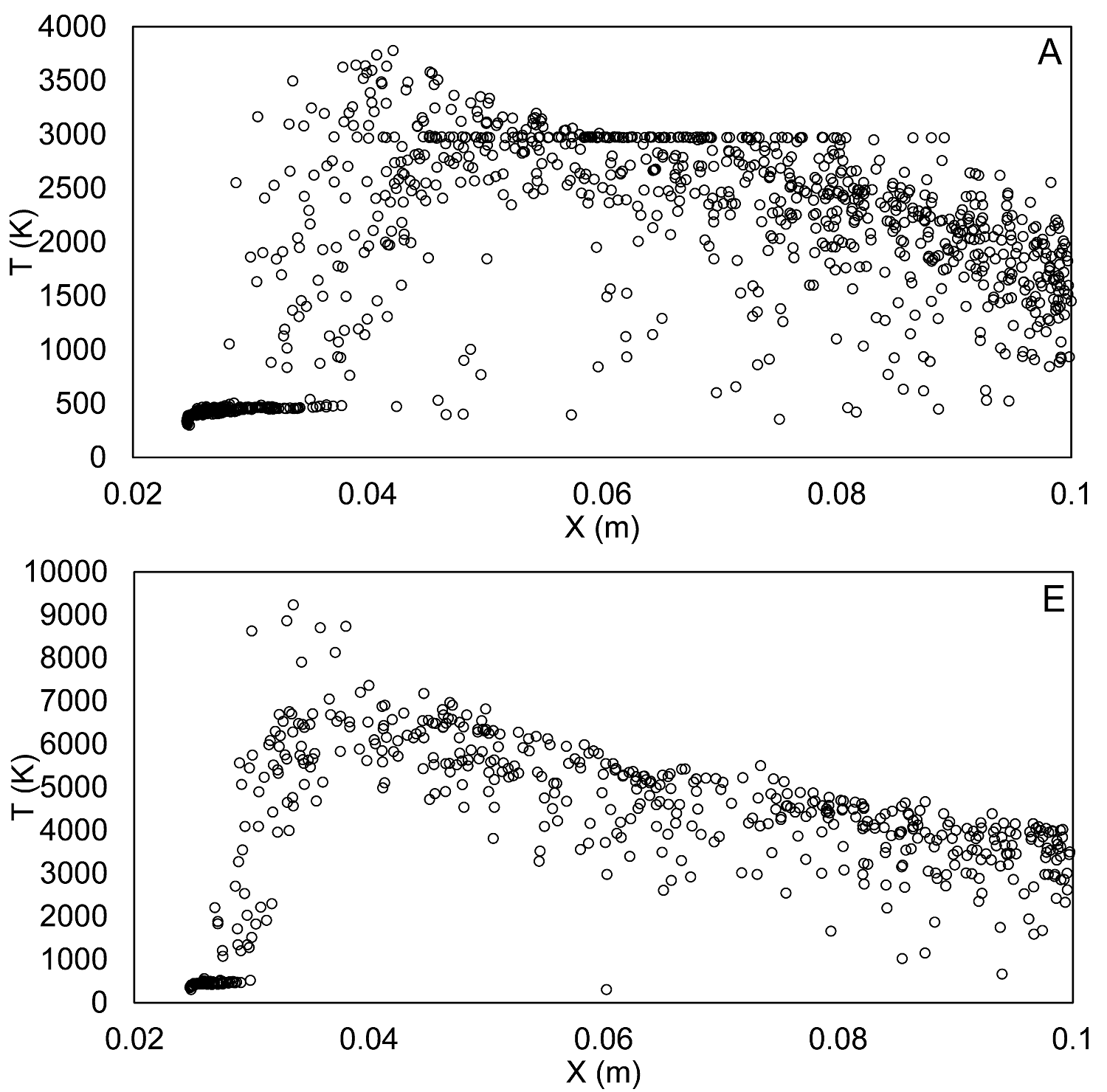}
\caption{Temperature of particles during flight for cases A and E.}
\label{fig:SPS_22}
\end{figure}

Temperature profile for the cases A and E have been compared in Fig. \ref{fig:SPS_22}. This snapshot of particles during flight shows that whether the particles have melted or not, still have their maximum temperature at 15 to 20mm from nozzle exit. After this point, the temperature starts to drop. This is mainly due to the fact that the plasma flow, as seen in Fig. \ref{fig:SPS_17}, has started to cool down close to this point. The rise in particle temperatures also lowers the rate of heat transfer. Particles now being at a higher temperatures will dissipate heat more rapidly through radiation. As be seen later, from the test cases here, more particles in cases D and E have temperature values higher than melting point of YSZ. This suggests the increase of power from 7kW to 14kW has a direct impact on better evaporation of liquid phase and melting of the solid YSZ particles. Temperature profiles of particles are nearly the same for cases D and E which indicates that the rise of power from 14kW to 21kW does not significantly increase the inflight maximum temperatures.

\subsection{Substrate Collections}

\begin{figure}[H]
        \centering
                \includegraphics[width = 6.5in]{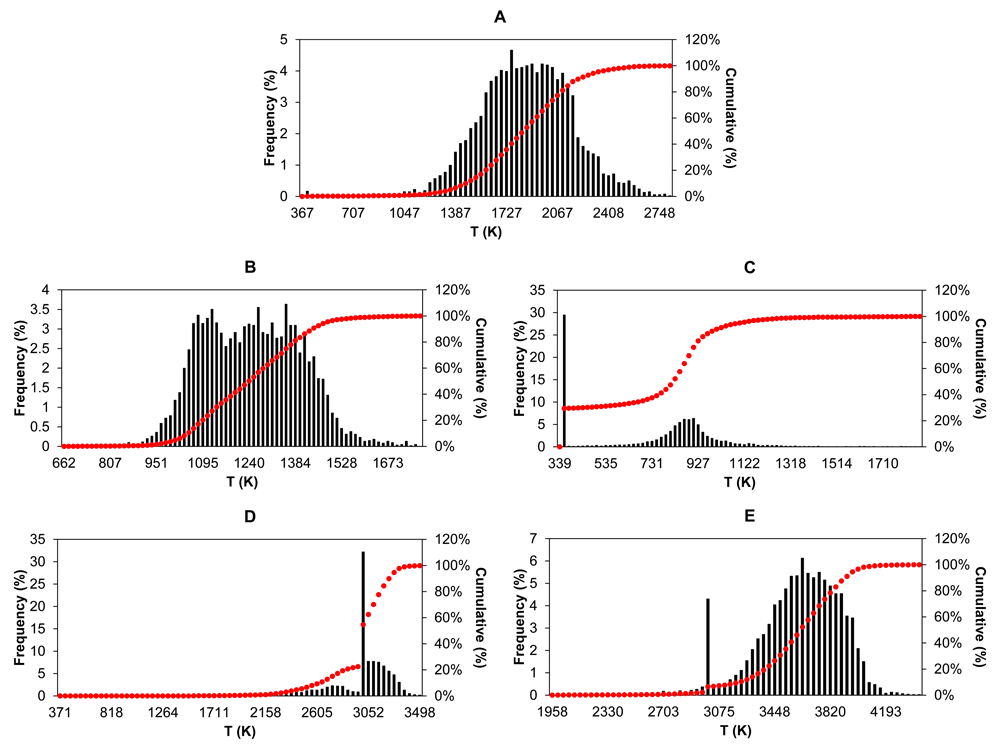}
\caption{Particle temperature distributions for test cases A-E upon impact on the substrate.}
\label{fig:SPS_23}
\end{figure}

Another important location for obtaining particle variables is on the substrate. Fully molten particles travelling at different speeds will generate different surface finishes. Moreover, under certain conditions, there is the possibility of having unmolten, semi molten, or even wet mixtures impacting the substrate. For the test cases here, a large substrate is placed 8cm downstream of nozzle exit. This surface will generate a stagnation flow pattern which helps in understanding of the particle interactions with the substrate. In previous studies, particles are typically collected in a region near the substrate. Here however, particles are only collected upon impact on the substrate. The rest of particles that come close to the wall but manage to escape are not included in here. It is also assumed that particles hitting the surface will adhere to it and do not rebounce.
The temperature distribution for the particles captured on the substrate for cases A-E are plotted in Fig. \ref{fig:SPS_23}. As the inlet flow rate of Ar-H2 changes from 35slpm for case A to 70slpm for case B and 140slpm for case C, the peak of the particle temperatures is shifted to lower temperature values. This shows that by keeping the torch power constant, the higher flow of gas has generated cooling effects which eventually has resulted in lower particle temperature. Particles travelling at higher speeds have less time for proper heat transfer. For these cases, the liquid content in some of the particles has not been fully evaporated by the time they reach the substrate. This is shown in the small peak below 500K, close to the evaporation temperature of water. As expected, the number of wet particles for case C is larger than cases A and B.

\begin{figure}[H]
        \centering
                \includegraphics[width = 6.5in]{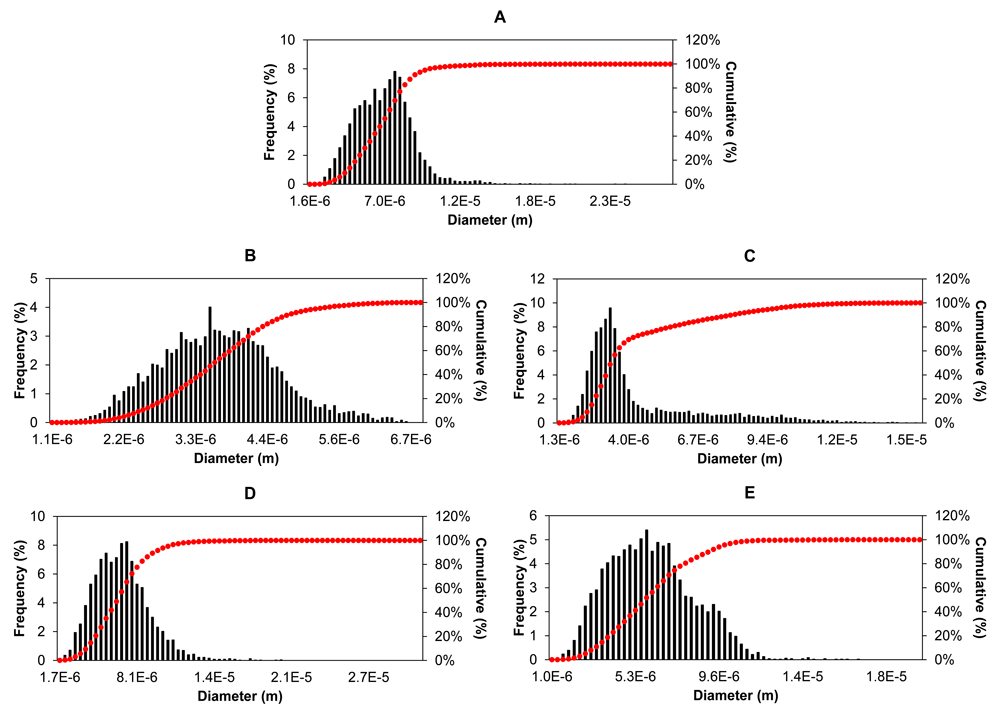}
\caption{Particle diameter distributions for test cases A-E upon impact on the substrate.}
\label{fig:SPS_24}
\end{figure}

Particles impacting the substrate for cases D and E have higher temperatures. For case D, most of the particles are semi-molten, explained by the sharp peak near the melting point of YSZ. This means that energy of the torch at 14kW was only enough to fully evaporate liquid phase and increase particle temperatures to the melting point. However, it was insufficient to help particles go through their melting process.
Fig. \ref{fig:SPS_24} shows particle diameter distributions for all test cases as they impact the substrate. These values are the final diameters of the particles after all the evaporations and breakups have taken place. Results show the increase of mass flow from case A to C shifts the peak to lower diameter sizes. The change from 35slpm to 70slpm seems to have a more significant effect compared to the change from 70slpm to 140slpm. The increase of the mass flow to 140slpm has only made the diameter range slightly narrower. Near the values of 70-140 slpm, most particles have diameters close to 3μm. The diameter for particles here are slightly larger than previously reported experimental values. This can be explained since the scattering and explosion of solid YSZ particles are not included in the assumptions. In reality, the solid particles inside the suspension do not form one sphere and can decrease the final size distribution of droplets.

\begin{figure}[H]
        \centering
                \includegraphics[width = 6in]{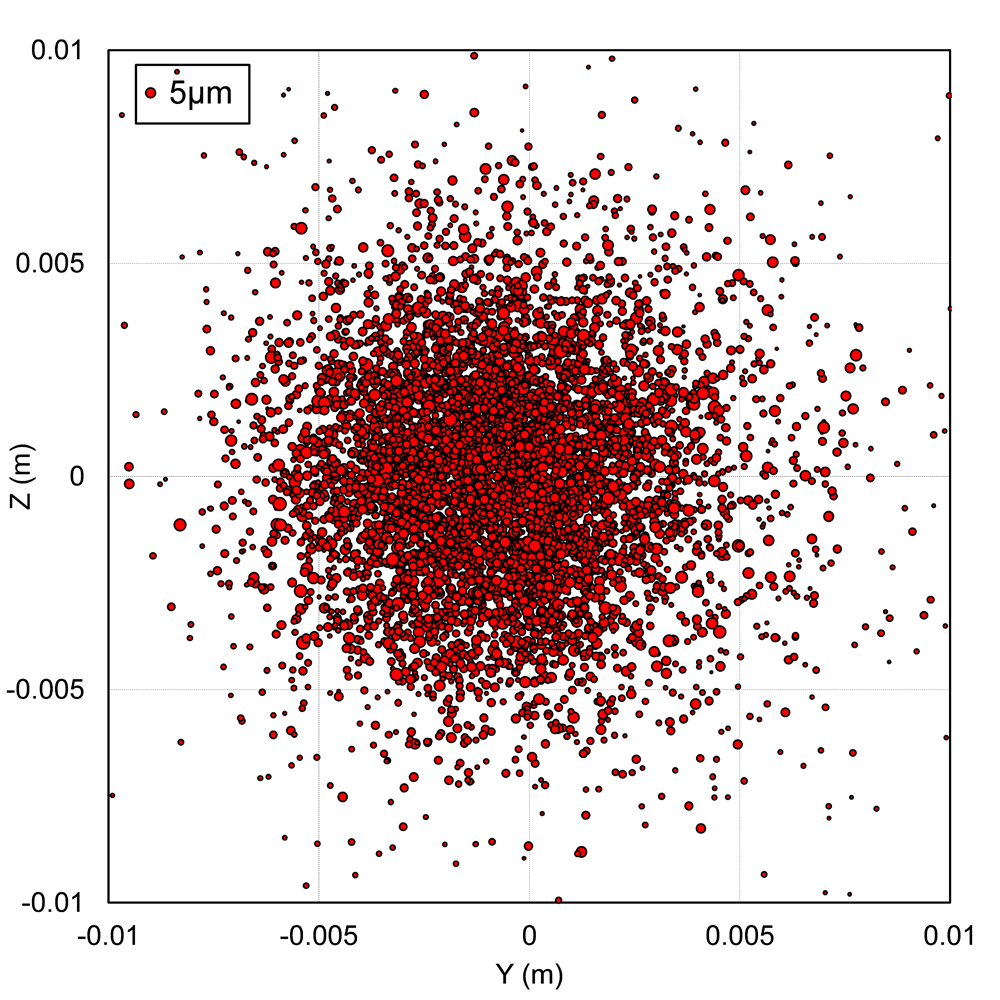}
\caption{Particles captured on the substrate 8cm downstream of torch for test case B collected over 3ms.}
\label{fig:SPS_25}
\end{figure}

Increasing the power from test case A to D and E also affects the final particle dimeters. Comparing the distribution of diameters between these three cases shows that the particles for all three cases nearly cover the same range of diameters. By increasing the power, however, the distributions have become less spread. For case A, for instance many particles reaching the substrate are still wet, carrying unevaporated water. These particles have gone through several stages of breakup due to shear force exerted by plasma flow. However, these suspensions have not managed to absorb sufficient heat to lose their moisture. Fig. \ref{fig:SPS_25} shows the particles from test case B that have been captured on the substrate over a certain time period. This figure shows that the torch footprint has slightly shifted to the left. This is explained by the fact that the suspension was injected from right to left.

\begin{figure}[H]
        \centering
                \includegraphics[width = 6.5in]{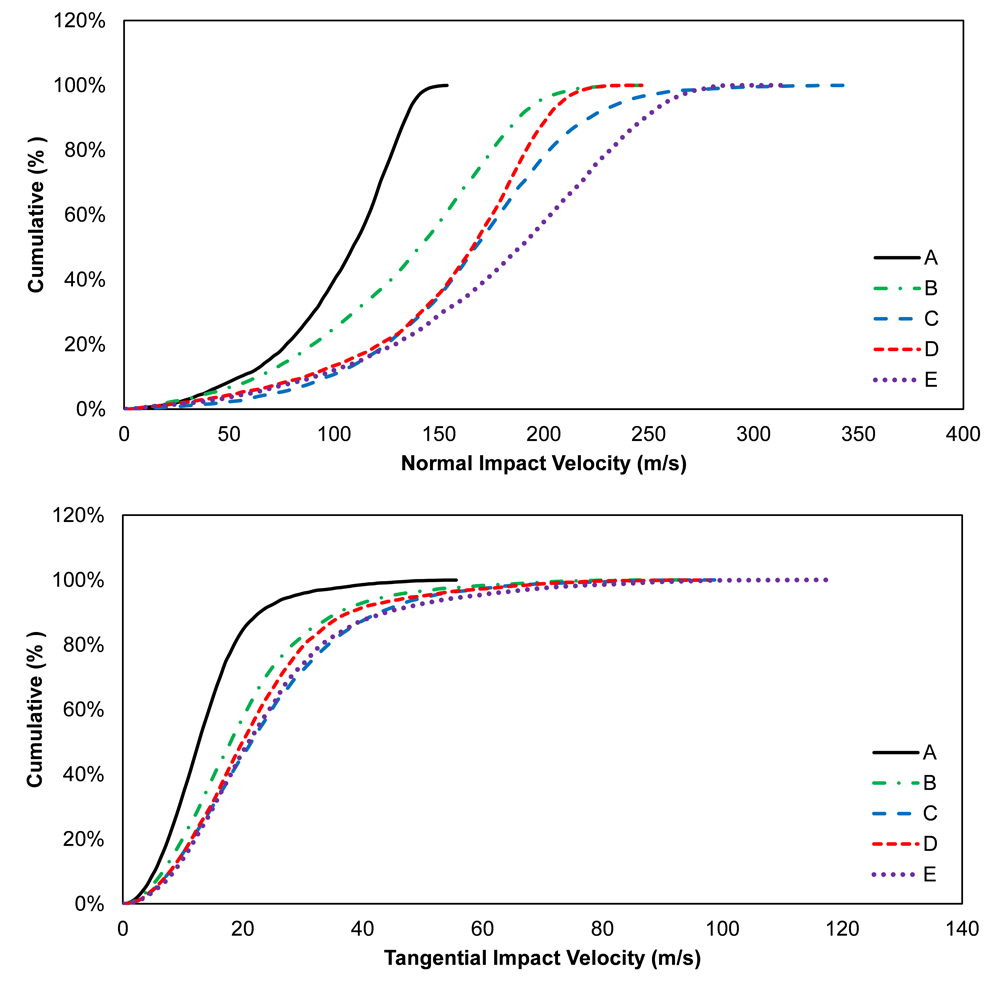}
\caption{Particle normal and tangential velocity distributions for test cases A-E upon impact on the substrate.}
\label{fig:SPS_26}
\end{figure}

The cumulative velocity distributions for particles upon impact on the substrate have been plotted in Fig. \ref{fig:SPS_26}. The velocity of particles has been divided into two components: one is the normal velocity which is the velocity component perpendicular to the substrate, and the other is the tangential velocity which is the velocity component parallel to the substrate. In an ideal situation, all momentum energy in the particles should be invested in a normal impact, making the tangential velocities zero. This is however impossible, as the stagnation flow pattern near the substrate will force particles to lose normal momentum. Case A here has lower tangential velocities. Increasing the power and mass flow rate have both led to higher velocities. The tangential velocities for cases B, C, and E have a similar pattern. The normal velocities here for all cases cover the same velocity ranges. Particles travelling close to the torch center line maintain higher velocities. As a typical substrate is smaller than the plane of study here, it can be predicted that most of the particles impacting the substrate will have the higher velocity ranges. However, for when the torch is not stationary, or the substrate of interest is large, the particles with smaller velocities will also impact the substrate. Different velocities upon impact will result in different splat sizes which will affect the coating microstructure. The normal velocity values follow nearly the same trend for all test cases. The line for test case B is higher than other cases at low velocities. Examining particle positions when impacting the substrate for case B reveals that particles at velocities higher than 25m/s have ended in a circle in the middle of the substrate. Particles with lower velocities however have covered a larger radius. Diameter measurements for both fast and slow particles are nearly the same. All particles with lower velocities also have lower temperature. This suggests that for this test case, the particles have been travelling close to the main plasma flow. These particles were unable to escape the main flow pattern, yet were not close enough to the centerline in order to receive enough heat. For the rest of test cases, it can be observed that the rise in power from 7kW to 21kW has clearly given particles higher impact velocities.

\section*{Conclusion}
Properties of suspension droplets is numerically studied during flight and upon impact on the substrate. Change of properties for suspension droplets during flight is taken into account here. Results clearly indicate inclusion of a proper viscosity model for the suspension, as it goes through large concentration changes, is vital in capturing accurate flight and breakup patterns. Moreover, the effect of different parameters on final deposition of particles and flight conditions is evaluated. Injection position, angle, and velocity, along with torch operating conditions such as flow rate and power are varied to create a range of test cases for better understanding of the suspension spraying process. These cases show that the final faith of the injected droplets is related to all parameters involved. Different parameters, however, have different impacts on the overall outcome. For injector operating conditions, test cases show that injection of suspension at high mass flow rates can result in a complete penetration through the plasma plume. It can also be concluded that under the conditions of the current study, the best SPS deposition rates are achieved when injection needle is placed close to the torch centerline and near 1cm from nozzle exit. Droplets also need to be injected at a low velocity. These injection conditions introduces the least amount of disturbances in torch flow pattern and allows more particles to be carried towards the substrate. For injection of water suspensions, it is clearly important to have enough power at the torch to make sure the cooling effects of evaporating water is recovered. Otherwise in many cases, the liquid content might not effectively evaporate and lead to the impact of wet mixtures on the substrate. Results here also indicate that increasing the torch power enhances particle qualities on the substrate better compared to increasing the inlet mass flow rate.

\bibliographystyle{unsrt}
\bibliography{References}

\end{document}